\documentclass[12pt]{article}
\pdfoutput=1 

\usepackage{geometry}
\usepackage[english]{babel}
\usepackage[utf8]{inputenc}
\usepackage{graphicx}
\usepackage{float}

\usepackage[colorinlistoftodos]{todonotes}
\usepackage{indentfirst}
\usepackage[titletoc]{appendix}

\usepackage{subcaption}
 \usepackage[backend=biber, style=authoryear,]{biblatex}
\appto{\bibsetup}{\raggedright}
\usepackage{csquotes}

\usepackage{amsmath}
\usepackage[mathscr]{euscript}
\usepackage{amsfonts}
\usepackage{amsmath}
\usepackage{amssymb}

\usepackage[linesnumbered,ruled,vlined]{algorithm2e}
\usepackage{setspace}

\usepackage{booktabs}
\usepackage{multirow}
\usepackage[labelfont=bf]{caption}

\usepackage{pdflscape,array,booktabs}
\usepackage{rotating}

\usepackage{hyperref}
\hypersetup{
    colorlinks,
    citecolor=black,
    filecolor=black,
    linkcolor=black,
    urlcolor=black
}

\title{Empirical Study of Market Impact Conditional on Order-Flow Imbalance}
\author{\\Anastasia Bugaenko\\
Supervisor: Dr Ankush Agarwal \\
Word Count: 14300}
\date{August 2019}

\begin{document}
\maketitle
\pagenumbering{gobble}

\begin{center}
A dissertation submitted in part requirement for the Master of Science in Quantitative Finance
\end{center}
{\let\thefootnote\relax\footnote{Adam Smith Business School, University of Glasgow}}
{\let\thefootnote\relax\footnote{Ana@symbiotica.ai}}
\thispagestyle{empty}

\clearpage
\renewcommand{\abstractname}{Acknowledgements}

\begin{abstract}
    \vspace{5mm}
    \centering I would like to express my gratitude to my family and my partner for their endless support, love and patience.  
\end{abstract}

\clearpage
\renewcommand{\abstractname}{Abstract}

\begin{abstract}
\vspace{5mm}
\noindent
In this research, we have empirically investigated the key drivers affecting liquidity in equity markets. We illustrated how theoretical models, such as Kyle's model, of agents' interplay in the financial markets, are aligned with the phenomena observed in publicly available trades and quotes data. Specifically, we confirmed that for small signed order-flows, the price impact grows linearly with increase in the order-flow imbalance. We have, further, implemented a machine learning algorithm to forecast market impact given a signed order-flow. Our findings suggest that machine learning models can be used in estimation of financial variables; and predictive accuracy of such learning algorithms can surpass the performance of traditional statistical approaches.

\vspace{5mm}
\noindent
Understanding the determinants of price impact is crucial for several reasons. From a theoretical stance, modelling the impact provides a statistical measure of liquidity. Practitioners adopt impact models as a pre-trade tool to estimate expected transaction costs and optimize the execution of their strategies. This further serves as a post-trade valuation benchmark as suboptimal execution can significantly deteriorate a portfolio performance.

\vspace{5mm}
\noindent
More broadly, the price impact reflects the balance of liquidity across markets. This is of central importance to regulators as it provides an all-encompassing explanation of the correlation between market design and systemic risk, enabling regulators to design more stable and efficient markets.

\end{abstract}
    
\vspace{5mm}

\vspace{5mm}
\noindent
\textbf{Note:} \textit{This copy of the research does not include the source code. Please contact the author for reference to the source code at – Email: ana@symbiotica.ai}

\clearpage
\tableofcontents

\clearpage
\listoftables
\addcontentsline{toc}{section}{\numberline{}List Of Tables}

\clearpage
\listoffigures
\addcontentsline{toc}{section}{\numberline{}List Of Figures}

\newpage
\pagenumbering{arabic}  
\setcounter{page}{1}
\section{Introduction}
\vspace{5mm}
A security marketplace broadly refers to any venue where buyers and sellers culminate to exchange resources, enabling prices to adapt to supply and demand (Bouchaud et al., 2018). Trading can take place in several possible ways; via broker-intermediated over-the-counter (OTC) deals, specialized broker-dealer networks, decentralised internal chat rooms where traders engage in bilateral transactions, amongst others.

\vspace{5mm}
\noindent
In traditional \textit{quote-driven markets}, all trading is enabled by designated market makers (MM or specialists liquidity providers) who quote their prices with corresponding volumes (the quantity to be bought/	sold), whilst other participants – market takers – submit their orders to either buy at quoted ask price or sell at the bid price posted by the market maker. In this respect, market makers offer indicative prices to the whole market. However, today, most modern markets operate electronically across multiple venues, and center around a continuous-time double-auction (where participants can simultaneously auction buy and sell orders) mechanism, using a visible limit order book (LOB). The LOB mechanism allows any participant to quote bid/ ask prices, and a transaction takes place whenever a buyer and a seller agree on the price. The London, New York (NYSE), Swiss, Tokyo Stock Exchanges, NASDAQ, Euronext, and other smaller markets operate using some kind of LOB. These cover a range of \textit{liquid} (traded in large volume) products including stocks, futures, and foreign exchange. Market participants in such venues can see the proposed prices, submit their own offers and execute trades by sending relevant messages to the LOB. Owing to technological developments, traders across the globe can access information about LOBs state in real-time and incorporate their observations when deciding on how to act. This transparency combined with low-latency, high liquidity and low trading costs of electronic exchanges appeals to many individual and institutional traders (Hautsch and Huang, 2011).

\vspace{5mm}
\noindent
The quality of a security market is often characterised by its \textit{liquidity}. Nevertheless, the term is not simple to define accurately, with precise definitions only existing in the context of particular models. Generally, liquidity is provided when counterparties enter into a firm commitment to trade. This ultimately results in an exchange of resources at a perceived free market \textit{fair price} (market clearing, as described by general equilibrium pricing). In this regard, the term captures the usual economic concept of price elasticity – in a highly liquid market (where many participants are willing to trade)  a small shift in supply (respectively demand) does not result in a large price change (Hasbrouck, 2007). Kyle (1985) more adequately describes liquidity by identifying three key properties of a liquid market: \textit{tightness} – “the cost of turning around a position over a short period of time”, \textit{depth} – “the size of an order-flow innovation required to change the prices by a given amount” or the available volume at the quoted price, and \textit{resilience} – “the speed with which prices recover from a random, uninformative shock”.

\vspace{5mm}
\noindent
Despite these simplistic yet elusive definitions, in the marketplace, liquidity is a complex variable with multiple unobservable facets, and often the main contributor to the non-stationarity of financial time series (amongst other variables, i.e., volatility). The difficulty in providing a more comprehensive definition of liquidity is exacerbated by the fact that academia has traditionally preferred to look at the world through the lens of a perfect, frictionless market with infinite liquidity at the market price. Nonetheless, the qualities associated with the word are sufficiently widely accepted and understood, making the term useful in practical discourse.

\vspace{5mm}
\noindent
In particular, practitioners discern market liquidity from that of funding liquidity. To capital market participants, liquidity generally refers to implicit or explicit \textit{transaction costs} (arising from limited market depth in the security), \textit{bid-ask spread} (i.e., quality spread – a difference in interest rates/ the difference in price at which one can buy or sell an asset) and \textit{price impact} (a change in market price that follows a trade). This is colloquially referred to as \textit{market liquidity}. Conversely, risk managers are often concerned with \textit{funding liquidity}.  This pertains to the ease at which a financial institution can raise funds/ capital to meet cash shortfalls (Acharya, 2006).

\vspace{5mm}
\subsection{Background}
\vspace{5mm}

\subsubsection{Liquidity Providers: The Modern Market-Maker}
\vspace{5mm}
\noindent
As outlined above, prior to the widespread adoption of LOBs, liquidity provision was traditionally designated to a small group of specialists. These specialists served as the exclusive source of liquidity for an entire market. This mechanism worked particularly well for quote-driven markets, granting these so-called MMs several privileges in exchange for immediate quotation and clearing services (i.e., ensuring settlement of transactions). To maintain efficiency under this market structure, dealers/ MMs must maintain undesirably large inventories (long position – assets that have been bought; short position – asset borrowed against a deposit known as collateral), accumulated whilst providing liquidity. This is problematic for MMs who typically aim to keep their net inventory as close to zero as possible, so as not to bear the risk of the assets’ price declining (Bouchaud et al., 2018).

\vspace{5mm}
\noindent
Alternatively, modern markets place no such restriction; in today’s electronic markets, all agents can act as MM by offering liquidity to other participants. This emerging complexity of electronic trading venues has intrinsically blurred the line between the usual distinction of liquidity provider (MM) and consumer. Nonetheless, to assist our discussion we adopt a more concrete distinction of the type of participants, as outlined in the works of  Cartea, Jaimungal and Penalva (2015) and Bouchaud et al. (2018):

\begin{enumerate}
  \item \textbf{Informed Traders} – attributed to sophisticated traders who profit from leveraging statistical \textit{information} (i.e., private signal or prediction) about the future price of an asset, which may not be fully reflected in the assets spot price
  
  \item \textbf{Uninformed Traders} – attributed to either unsophisticated traders with no access to (or inability to correctly/ efficient process) information, or market participants who are driven by economic fundamentals outside of the exchange. These traders are often labelled \textit{noise} traders as a large fraction of their trades arise from portfolio management and risk-return trade-offs that carry very little short-term price information
  
  \item \textbf{Market makers (MMs)} – attributed to (provisionally) uninformed professional traders who profit from facilitating the exchange of a particular security and exploiting their skills in executing trades
\end{enumerate}

\vspace{5mm}
\noindent
Clearly, the notion of \textit{information} fundamentally underpins our classification of each agent and defines their ability to accurately forecast price changes (Bouchaud et al., 2018). Considering the interactions and tensions amidst these groups provides useful insights into the origins of many interesting observed phenomena in modern financial markets.

\vspace{5mm}
\subsubsection{Asymmetric Information and Adverse Selection}

\vspace{5mm}
\noindent
The rate at which information is incorporated/ reflected in prices underpins the degree of \textit{efficiency} in the market. In this regards, financial markets are not generally classified purely at two extremes (efficient or inefficient) but have been shown to exhibit various degrees of efficiency (McMillan et al., 2011). In this view, market efficiency is observed as a continuum between extremes of completely efficient, at one end, and inefficient at the other. This is consistent with widespread empirical observations (see  Finnerty (1976) and  Seyhun (1986)), where the strong form efficiency has been shown not to hold in light of private information.

\vspace{5mm}
\noindent
As private information can consist of signals about the terminal value of the security, information asymmetry is of fundamental importance to MMs (who often trade with highly informed participants) and is the prevailing consideration of our study. Whereas most small trades contain relatively little information and are thus innocuous for MMs providing liquidity; larger orders could be interpreted as stronger signals of an information advantage stemming from better predictive models.

\vspace{5mm}
\noindent
An imperative consequence of such \textit{informed order-flows} (trends in the direction of trading arising from more informed participants) is the resulting inventory imbalance, where MMs are forced to accumulate larger net positions in the short-run – i.e., MM receives many more buy orders than sell, with a high probability of being on the wrong side of the trade. This is known as \textit{adverse selection} and may cause MMs huge losses as they are “picked-off” by more informed traders when making binding quotes (Hasbrouck, 2007).

\vspace{5mm}
\noindent
To compensate for this information asymmetry (therefore mitigating the risk of being adversely selected), MMs choose how much liquidity to reveal and look to efficiently process any new piece of information by updating their bid/ ask quotes in response to the order-flow imbalance. Such market friction results in Mean Field Games, where MMs adjusts their bid/ ask prices as more informed liquidity takers submit large trades. This leads to a worse execution price for the informed trader – the so-called \textit{market} or \textit{price impact}. Consequently, informed agents must selectively take liquidity using \textit{optimal execution} strategies (i.e., split their large orders across time to match the liquidity volume revealed by MM) as described in the work of  Almgren and Chriss (2001), see Appendix B.1.

\clearpage
\vspace{5mm}
\subsection{Motivation}
\vspace{5mm}
\noindent
Following the wake of the 2007 global credit crisis, there has been a myriad of regulations requiring institutional investors (both on the buy-side and sell-side) to meet several liquidity related policies (see Table~\ref{table:regulation}). This stems from the general perceived reduction in the quality of liquidity across asset classes as per the report produced by Bloomberg (2016).

\vspace{5mm}
\begin{table}[hbt!]
\centering
\begin{tabular}{c c c }
\hline\hline 

 & Buy Side                          & Sell Side  \\ \midrule
 & Prudent Valuation                 &  MIFID II            \\
 & RRP                               &  SEC (22E-4)            \\
 & ILAAP                             &  AIFMD             \\
 & Basel 3 (LCR)                     &  UCITS              \\
 & FRTB (Basel 4)                    &  FORM PF               \\

\hline\hline 
\end{tabular}
\caption{Financial Liquidity Regulations}
\label{table:regulation}
\end{table}


\vspace{5mm}
\noindent
Liquidity risk is of special importance to practitioners because it might cause a bank to fail despite no trading losses (Murphy, 2008). This risk pertains to the firms’ ability to meet cash demands. These demands might be either known in advance, such as coupon payments; or unexpected, such as the early exercise of options or the need to liquidate portfolios of large positions. Therefore, inadequate funding and market liquidity may impair the firms' ability to meet their payment obligations. 

\vspace{5mm}
\noindent
Moreover, excess transaction costs arising from liquidity concerns are important factors in determining investment firms’ performance. These costs can become very high, reducing any trading profits. According to Jean-Philippe Bouchaud from Capital Fund Management, nearly two-thirds of trading profits can be lost because of market impact costs (Day, 2017). Whilst explicit transaction costs can be accounted for, the implicit costs (such as market impact) cannot be estimated directly but can be approximated by measuring liquidity and minimized by adopting an optimal trading strategy.

\vspace{5mm}
\noindent
Within the microstructure of financial markets, an \textit{optimal liquidation/ acquisition} strategy delivers the minimum market impact for a particular order size and time horizon (the urgency at which an asset is to be bought/ sold). In this respect, Kyle and Obizhaeva (2018) define market impact as the expected adverse price movement from a pre-trade benchmark (the decision/ fair price for which a trader wishes to purchase an asset), upon execution. Consequently, accurate measurement of market impact is essential, possibly blurring the line between a profitable and unprofitable strategy net of such transaction costs.

\vspace{5mm}
\noindent
However, the effect of a firm's own trading activity on the market prices is notoriously difficult to model as there is no standard formula that applies to every financial asset or trading venue.  Deriving such formula is a challenging task due to the lack of trade activity and data in various asset classes. For instance, investment-grade fixed income securities are traded in quote driven over-the-counter markets (OTC), with no transaction visibility, whereas large common stocks are often found in more liquid order driven electronic exchanges. Hence, the functional formula for the market impact would vary according to assets characteristics and trading pattern.  

\vspace{5mm}
\noindent
Market microstructure literature has discussed a number of market impact/ cost functions, with theoretical studies arriving at a model of linear functional form, where price impact is said to be proportional to the volume of security traded in the market. On the other hand, an overwhelming number of practitioners have purported a square root model, which suggests a marginal price impact diminishes as the trade volume increases (Kyle and Obizhaeva, 2018). Despite the presence of some empirical evidence for the square root model of market impact, both practitioners and academics agree that the model is not exact and is not aligned with the theoretical research (Bouchaud, 2009). 

\vspace{5mm}
\noindent
The purpose of this study, therefore, is to conduct a robust empirical analysis of the market impact functional form and validate it against existing models. The novelty of our methodology lies in the advanced statistical tools adopted. Specifically, the study will examine the application of machine learning techniques to the derivation of a market cost function. Unlike traditional regression analysis that suffers from limitations such as “\textit{curse of dimensionality}” (the model becomes mathematically intractable when dealing with a large number of explanatory variables), machine learning (ML) has been proven to provide robust results in many higher-dimensional financial applications. This is because of its ability to fit and predict using complex data sets (Park, Lee and Son, 2016). With the increasing availability of high-frequency market trading data, we are now at an acute juncture where we can begin to conduct meaningful studies of the relationship between order flow, liquidity and price impact in order-driven markets. In doing so we hope to facilitate a better understanding of market impact function given the gap between current empirical findings and the theory.   

\vspace{5mm}
\noindent
Being able to model market impact more accurately is essential for a better understanding of how trades affect prices and how to quantify the degree of this impact as well as it’s dynamics (Guéant, 2016). This knowledge of the price formation process would empower both practitioners and academics to arrive at models that better depict observed market behaviour, further contributing to the efficiency and stability of modern market microstructure.

\vspace{5mm}
\subsection{Research Objective}
\vspace{5mm}
\noindent
The focus of this research is to derive a functional form of market impact using parametric ML algorithm. To achieve this, a detailed investigation of the key drivers affecting liquidity is required; with an emphasis on observing the consequences of executing large orders by exploiting data from a stock exchange.

\vspace{5mm}
\noindent
A series of experiments will be carried out using the scientific method to statistically reconstruct the dynamics of NASDAQ Limit Order Book (LOB). LOBs contain detailed information about the interplay between liquidity providers (i.e., market makers) and liquidity takers. This permits us to select microstructure features (explanatory variables) that underpin price impact, and thus, need to be included in its function definition. These features will then serve as input features to the ML algorithm.

\vspace{5mm}
\noindent
The work will be conducted in a controlled environment, examining liquid stocks. This allows for a vast and rich data set that can facilitate precise and robust numeric results. As parametric models are often prone to overfitting (thus, bad forecasting), we look to reduce both bias and variance errors by conducting cross-validation of the derived model. The latter involves dividing the training data set in random parts and fitting the model on each partition (known as out-of-sample testing).

\noindent
\textbf{To summarise, the aims of this study are:}

\begin{enumerate}
  \item Investigate key drivers affecting liquidity in equities markets
  \item Derive a functional form of market impact using machine learning algorithm 
  \item Compare machine learning predictive performance against traditional statistical models using cross validation
\end{enumerate}

\vspace{5mm}
\subsection{Thesis Structure}

\vspace{5mm}
\noindent
The structure of this thesis is organised as follows:
\begin{itemize}
    \item Chapter 2 - \textit{Background and Literature Review} is a review of key terms and introduction to the research environment.
    
    \item Chapter 3 - \textit{Data and Research Methodology} describes the dataset and the approach adopted throughout the study.
    
    \item Chapter 4 - \textit{Empirical Study} illustrates the outcomes of our experiments; and discusses the implications of the findings.
    
    \item Chapter 5 - \textit{Conclusion and Future Work} is a summary of our key results and suggestions for future works.
    
\end{itemize}


\section{Literature Review}
\vspace{5mm}
No respectable model exists without an appropriate understanding of the system rules and challenges faced by domain practitioners, as well as empirical facts. To facilitate a comprehensive study of the functional form of market impact, we must first consider several key concepts present in the \textit{Market Microstructure} literature.

\vspace{5mm}
\noindent
Market microstructure forms a long and rich history of differing viewpoints, with academics (economist, physicists, and mathematicians) and practitioners (regulatory policymakers and investors) typically residing at two distinct ends of the spectrum. As we will discuss, all such perspectives have their confines and intersect. Developing a coherent understanding of these themes is a long and complex endeavour. This chapter serves to situate these issues within the current research anatomy.

\vspace{5mm}
\subsection{A Brief Primer on Market Microstructure}
\vspace{5mm}
\noindent
The microstructure of a market is characterised by the interactions of the kinds of participants, and rules governed by regulators. These rules focus on minimizing any friction arising at the level of trading venues, as well as how the exchange of assets takes place in very specific settings.

\vspace{5mm}
\noindent
The term market microstructure was first coined by Garman (1976), in the paper of the same title, where he describes the moment-to-moment trading activities in asset markets. The field has since emerged as an effervescent research area of prominent importance. A substantial number of changes have occurred since the expressions first usage. For example, the price formation process has been impacted by the fragmentation of markets in major financial hubs such as the US and Europe (e.g., introduction of \textit{Dark Pools} – alternative trading systems with no visible liquidity, for which market activities take place away from public exchanges), no doubt due to the abundance of technological advances (i.e., automation of trading and the development of execution algorithms). These modern market designs have prompted new questions for modelers.

\vspace{5mm}
\noindent
However, information remains a key dimension at the heart of the prevailing microstructure studies. The first iterations of models embracing this notion of information were developed during the last quarter of the 20th century. Economists such as Kyle (1985), provided an in-depth analysis of how information is conveyed into prices; and the impact of asymmetric information on liquidity in general. This notion that “market prices are an efficient way of transmitting the information required to arrive at a Pareto optimal allocation of resources” (Grossman, 1976) – is a natural emerging property of microstructure studies that aim to identify how different trading conditions and rules promote, or hinder, price efficiency. 
That is, many classical (static and dynamic) microstructure models describe the process by which new information comes to be reflected in prices. This transmission of information into transactions and prices is deeply related to market impact, provision of liquidity and determinants of the bid-ask spread. These topics were often the focus of much of the earlier academic literature.

\vspace{5mm}
\noindent
The first academic papers focussing on optimal execution were those of Bertsimas and Lo (1998), Almgren and Chriss (1999) and Almgren and Chriss (2001), with interest in the subject only truly proliferating beyond 2000 (Guéant, 2016). Models incorporating the use of \textit{limit orders} (visible orders resting in the LOB) and dark pools soon followed. Appendix B.2 describes earlier LOB models and their evolution.

\vspace{5mm}
\noindent
These new models featured more complex variables such as trading volatility and involve coefficients that need to be estimated using \textit{high-frequency} datasets (time series of market data observed at extremely fine scales, i.e., milliseconds). Many statisticians are now acutely involved in the study of market microstructure, bringing with them advanced methods based on stochastic calculus that allow for better estimation of parameters given the data. More specifically, there are several important pieces of literature on high-frequency liquidity provision. This began in 2008 with the publication of Avellaneda and Stoikov (2008) who presented a model of market dynamics, comprising of a complex partial differential equation (PDE) that was solved by Guéant, Lehalle and Fernandez Tapia (2013).

\vspace{5mm}
\noindent
Today, quantitative research on market microstructures is more concerned with the importance of pre- and post- trade transparency, the optimal tick size, the role of alternative trading venues, clearing and settlement of standardized products, amongst others.

\vspace{5mm}
\noindent
Nonetheless, there remains room for improvements when it comes to more realistic dynamic market models that better depict widely observed, but still poorly understood micro- and macro- structure phenomena (Bouchaud et al., 2018). To examine this relationship between market dynamics and some exogenous variables such as volume and order-flow imbalance, we must first review the properties of prominent models of market impact.

\vspace{5mm}
\subsection{Market Impact}
\vspace{5mm}
\noindent
As we have deliberated, the notion of liquidity in financial markets is an elusive concept. However, from a practical stance, one of its most important metrics is the response of price as a function of \textit{order-flow imbalance} (i.e., excess volume with respect to the order sign). This response is known as market impact. 

\vspace{5mm}
\noindent
In much of the literature, there are three distinct strands of interpretation for the cause of market impact, which reflects the great divide between efficient market enthusiasts and sceptics: 

\begin{enumerate}

    \item The Efficient Market Hypothesis (EMH) posits that all available information is reflected in prices as rational agents immediately arbitrage away any deviation from the fair price. In the efficient market framework, rational agents who believe that asset prices are always close to their fundamental value “successfully forecast short-term price movement”. This can result in a measurable correlation between trade sign and subsequent price change (Bouchaud, Farmer and Lillo, 2009). under this interpretation, as emphasised by Hasbrouck (2007), “…orders do not \textit{impact prices}. It is more accurate to state that orders \textit{forecast prices}”, thus, noise-induced trades carrying no information yield no long-term price impact as prices would otherwise deviate from their fundamental value.
  
    \item The second picture reinforces the first in that market impact is the apparatus by which prices adapt to new information as illustrated in the aforementioned Glosten and Milgrom (1985) and Kyle (1985) models; therefore permitting information about the fundamental value of the asset to be incorporated into prices. 
  
    \item The third perspective resonates with that of the efficient market sceptic. Here, in the absence of fundamental price nor private information, zero-intelligent models describe how prices impact is a reaction to order-flow imbalance. In Farmer, Patelli and Zovko (2005) Santa Fe model, they delineate a completely stochastic order-flow process by which the act of trading itself is tautologically seen as the physical medium statistically interpreted as price impact.
  
\end{enumerate}

\noindent
Though all three interpretations result in a positive correlation between trade signs, volume and price impact (response function), they are conceptually very distinct. In the first two pictures, trades reveal private information about the fundamental value of assets, resulting in a price \textit{discovery process}. In the latter mechanical interpretation, one should remain agnostic of the informational content of trades and should instead speak of \textit{price formation}.

\vspace{5mm}
\noindent
Trading impacts prices – this is an undisputable empirical observation. However, the interpretation of this impact is still widely debated; whether prices are formed or discovered remains a topic of discussion with no clear consensus at this stage. But because of the unclear distinction between true information and noise, one can assume reality lies somewhere between all three extremes (Bouchaud et al., 2018).

\vspace{5mm}
\noindent
The concepts of adverse selection and market impact are often captured by \textit{asymmetric information} models describing why liquidity providers actions should depend on the behaviour of other market participants. A key early reference on the subject is the seminal Kyle (1985), which provides an elegant explanation of how impact arises from liquidity providers fears of adverse selection when trading against highly informed traders. Albeit not very realistic, the model is a concrete illustration of how private information comes to be reflected in the price of an asset. Moreover, economic models further provide important insights into the challenges faced by MM. In this regard, Glosten and Milgrom (1985) work demonstrate how the bid-ask spread must compensate MM for adverse selection when trading in a competitive market. 

\vspace{5mm}
\subsubsection{Price Impact Models}
\noindent
The market microstructure literature has studied a diverse range of market impact models. However, despite many decades of theoretical and empirical research, there remains a vast number of open questions regarding its functional form (Kyle and Obizhaeva, 2018). This is in part due to the varying characteristics of assets as dictated by their market microstructure. Moreover, as outlined in Almgren et al. (2005), there are different classes of market impact that must be distinguished. Before presenting our empirical investigation, we consider some heuristic models of market price dynamics. These models intuitively encapsulate the strategic considerations of market participants by examining the delicate balance in liquidity maintained by ongoing competition between rational agents. 

\vspace{5mm}
\clearpage
\noindent
\textbf{Linear and Permanent Impact: The Kyle Model}

\vspace{5mm}
\noindent
Kyle (1985) proposed a classic toy model which seeks to shed light on the mechanisms by which private information is gradually propagated into prices in an efficient market. The model was further extended in (Kyle (1989) to account for the case of several competing informed traders. Kyle’s original model assumes the simple case of a normal distributed random variable (Kyle’s Lambda $\EuScript{\lambda}$) and derives a single statistical measure of the impact that is both linear in traded volume (order size) and permanent in time. In this framework, market dynamics are juxtaposed as a contest between an inside trader (who holds unique information regarding the fair price of an asset) and a noise trader (who submits random orders in the absence of actual knowledge) whom submit orders that are cleared by a Market Maker (MM) at every time step $\Delta t$. In the model, the price adjustment rule $\Delta p$ of the MM must be linear in the total signed volume $\varepsilon v$, i.e., 

\begin{eqnarray}
\Delta p = \lambda \varepsilon V
\end{eqnarray}

\vspace{5mm}
\noindent
This gauges the market impact of a trade as a consequence of volume flow imbalance, where $\lambda$ is a measure of impact and is inversely proportional to the liquidity of the market. The consecutive price adjustment is further \textit{permanent}, i.e., the price change between time $t =0 $ and $t = T = N \Delta t$ is:

\begin{eqnarray}
 p_t = p_0 + \sum_{n=0}^{N-1} p_n = p_0 + \lambda \sum_{n=0}^{N-1} \varepsilon_n V_n
\end{eqnarray}

\vspace{5mm}
\noindent
Formula 2 assumes that the impact $\lambda_n V_n $ of trades in the nth time interval persists unconditionally unabated up to time $T$. It is clear that the signs of the trade must be serially uncorrelated if the price is to follow a stochastic path. Within the model’s framework, the trading schedule of the informed insider is precisely such that $\varepsilon_n$ are uncorrelated (Kyle, 1985). Conversely, data from the real markets reveal autocorrelation in the signs of the traded volumes over prolonged time frames as empirically observed by Bouchaud et al. (2003).

\vspace{5mm}
\noindent
In summary, Kyle’s model elicits some elegant deep truths about how markets function, but also fails to capture the essence of important empirical properties of real markets. In the model, the mechanism by which information comes to be reflected in prices (price impact) is due to the MM’s attempt to predict the informational content comprised in the order-flow and adjusts their prices accordingly. Additionally, price impact is said to be linear (i.e., price changes are proportional to the order flow imbalance) and permanent (i.e., there is no decay of impact exhibited). In the context of LOBs, linear impact is a generic consequence of a finite liquidity density  (buy/ sell orders) in the vicinity of the price, and permanent impact is the outcome of liquidity immediately refilling the subsequent consumption gap caused by the market order, as demonstrated by Obizhaeva and Wang (2013). Because of impact, informed traders restrict their trading volume to optimise returns. Thus, despite private information, the amount of profit that can be made by exploiting this insider knowledge is limited.

\vspace{5mm}
\vspace{5mm}
\noindent
\textbf{Concave Transient Impact: The Square Root Model}

\vspace{5mm}
\noindent
Kyle’s original model assumes a linear dependency of impact on traded volume, but this requires a variety of idealised assumptions that may be violated in real markets (Zarinell et al., 2015). A key finding is that impact is not only mechanical, but also dynamic, meaning it cannot simply be described by revealed supply or demand of a visible LOB (Weber and Rosenow, 2005). The impact is rather related to the \textit{latent underlying liquidity} – hidden supply and demand, not reflected in the LOB. This stems from the fact that even \textit{highly liquid} markets only offer very small volumes of liquidity for immediate execution (see Appendix B.1). Consequently, trades must be fragmented creating long memory in the sign of the order-flow as private information is slowly incorporated into prices. This, however, is incompatible with the permanent impact (as described by (Kyle (1985)), which would otherwise lead to trends, i.e., strong autocorrelated price changes (Bouchaud, 2009).

\vspace{5mm}
\noindent
As an alternative, the square root model of market impact was first proposed by  Torre (1997) based on empirical regularities observed by Loeb (1983). Empirical studies have since ubiquitously found market impact to indeed be a non-linear concave function in the size of \textit{meta-orders} (large orders fragmented into a number of smaller market orders) and fading in time (Bouchaud, 2009). This concave nature of market impact is universally observed over several heterogeneous datasets in terms of markets (including equities, FX, futures, bitcoins and even OTC credit markets as reported by Eisler, Bouchaud and Kockelkoren (2012), epochs and execution styles. On this note, it is worth mentioning that some studies report empirical deviations from the square-root law, e.g., Almgren et al. (2005) and Zarinell et al. (2015).

\vspace{5mm}
\noindent
Nonetheless, in all former cases, the square root model is consistently simple and empirically realistic. Let $G$ denote the percentage cost of executing a meta-order of size $Q$ shares of a stock with price $P$, expressed as a fraction of the value of the trade $\left|P\ Q\right|$. Let $\sigma$ denote the assets return (daily) volatility, and let $V$ denote the assets (average) daily traded volume in shares per day. The square root law of market impact is thus described by the relationship

\begin{eqnarray}
G = g(\sigma, P, V; Q) \sim   \sigma\left(\frac{\left|Q\right|}{V}\right) ^\frac{1}{2}\
\end{eqnarray}

\vspace{5mm}
\noindent
where $g(\cdot)$ defines the functional form for the model and the notation "$\sim$" means “is proportional to”. The market impact $G$ is a dimensionless (absolute) quantity, which is the same regardless of the units of measurement for impact $Q,\ V,$ and $\sigma$.

\vspace{5mm}
\noindent
Even though model (3) seems a reasonable indication of transaction costs, there still no consensus on whether or not price impact is indeed described by the square root function. The exponent varies within a range of 0.4 to 0.7. For example, empirical findings by Almgren et al. (2005) and Kyle and Obizhaeva (2016) find an exponent closer to 0.6, but average observations suggest a power closer to  0.5  (square root). We note that the only conditional variable is the total traded volume $Q$. This is surprising as it implies that the time taken to completely liquidate (respectively acquire) and execution path are not important factors in determining market impact. Nevertheless, real data shows that impact does depend on such dynamics, thus, \textit{power law}, as often referenced, should, therefore, be seen as a good first-order-approximation. Whilst this result is in stark contrast with classical economic literature (which asserts linear impact (Kyle, 1985)), it is perfectly in accordance with the fact that instantaneous observed liquidity is limited in real markets further indicating that markets may be inherently fragile.

\vspace{5mm}
\noindent
Our empirical study aims to quantify how liquidity providers react to the arrival of market orders. The following section provides the foundation of the statistical approaches employed in this research. By outlining the analysis methodology and exploring more recent developments in the area of data analytics, we hope to facilitate a practical solution for accurately measuring order-flow, therefore, providing clarity on many obscurities surrounding price impact.

\section{Data and Research Methodology}
\vspace{5mm}

\subsection{Electronic Markets}

\vspace{5mm}
\subsubsection{Limit Orderbook (LOB) Trading}

\vspace{5mm}
\noindent
Digital markets, generally, facilitate trade by automatically matching those wanting to sell with those wanting to buy. Market participants express their willingness to trade a specific quantity at a specific price by submitting orders to the exchange. At a high level, all orders are classified by their type as Market Orders (MO) or Limit Orders (LO). MOs indicate an immediate need to execute the trade. LOs, on the other hand, are known as passive orders because these usually do not result in an instant execution. LOs are often submitted at a price worse than the current market price, and therefore, have to wait until either a new order arrives that matches LO price or the LO is withdrawn. All active (non-cancelled) LOs are placed in a queue according to their corresponding price. This \textit{order queue} is managed within the LOB, whilst mapping of orders is conducted by the matching engine.

\vspace{5mm}
\noindent
LOB is defined on a fixed discrete grid of prices where submitted LOs are recorded in separate price level queues. Figure 1 illustrates a sample snapshot of LOB: vertical blue and red bars represent queues of LOs to buy and sell respectively. The length of each queue is defined by the number and sizes of orders that have been submitted at that price, but not yet matched for execution. The scale of price and volume axis is defined by LOB resolution parameters – tick and lot sizes respectively. The tick size is the smallest possible change in the price of an asset. In other words, the tick size defines the precision of the quoted price. On the other hand, the lot size defines the smallest amount (expressed as the number of shares) of security that can be traded within the LOB.

\vspace{5mm}
\noindent
When a new buy or sell LO comes in, it is added to the end of the corresponding price queue – on the top of previous LOs at that level, see Figure 1. The difference between the best (lowest) ask and the best (highest) bid prices is known as the spread; while the arithmetic average of these best quotes is called the \textit{mid-price}. Mid-price is often used to describe the LOB and its dynamics (as opposed to looking at individual behaviour of the bid and ask prices). Therefore, when examining the market impact of orders, we are interested in how mid-price had changed in response to the execution of MOs. Note that in some cases, it is more appropriate to examine \textit{micro-price} – the weighted (by inverse volume) average of the bid and the ask. It comes to be more useful when the imbalance of the orderbook is used for prediction of the sign of future price changes as emphasised in Cartea, Jaimungal and Penalva (2015) and Bouchaud et al. (2018). In this study, however, we use mainly the mid-price.

\vspace{5mm}
\vspace{5mm}
\begin{figure}[hbt!]
    \centering
    \includegraphics[height = 7cm, width=9cm]{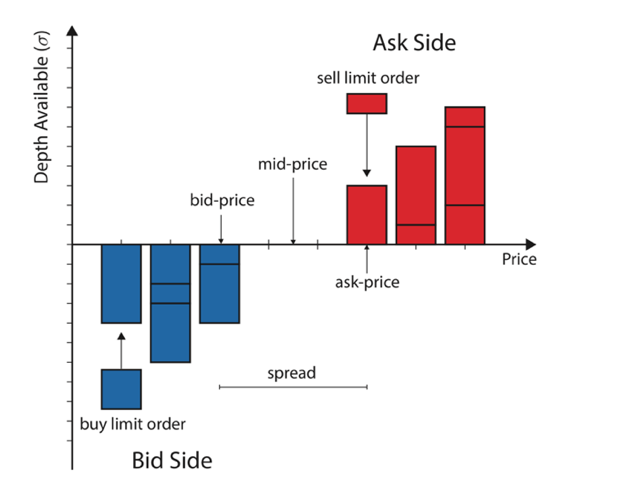}
    \caption{Illustration of LOB dynamics (Bonart and Gould, 2017).}
    \label{fig:lobDynamics}
\end{figure}

\vspace{5mm}
\noindent
Matching engine, or matching algorithm, is a well-defined procedure on how to orders are select and execute. Predominantly, in electronic markets, the algorithm tries to map MOs first – if two MOs match then these are executed immediately. If a new MO does not have an opposite side matching MO at the time of submission, it is executed against LOs in a price-time priority order. That is, an incoming MO is first mapped to the oldest LO at the opposite side best price. Then, if the quantity demanded by MO is not fulfilled, it is executed against earlier LOs (still at the best available price). It is interesting to note that not all matching engines abide by a price-time priority queuing mechanism. Alternative matching engines such as the \textit{prorata} rules are often found in alternative market structures (e.g., in money markets). Under the prorata model, there is no explicit time-priority rule. MOs are instead matched against LOs posted at the best opposite price – proportionate to the quantity posted. Moreover, there are other markets that combine the two approaches, adopting both time-priority and pro-rate (i.e., Futures).

\vspace{5mm}
\noindent
Traditionally, if the full size of MO cannot be fulfilled by LOs resting at the best price, the matching algorithm would try to execute the rest of MO against LOs at the second, third and so on best prices; until the full quantity of LO is filled (walking the book). However, modern financial markets implement alternative procedures for handling situations where there is not enough liquidity at the best price to which the entire MO can be matched. For example, in the US, trading venues are obliged by regulators to provide participants with the best possible price for the asset. This means that, depending on the specific order type, the exchange may re-route the remaining units of an unfulfilled order to alternative venue displaying the same best quote.  While re-routing can certainly affect liquidity (thus price impact) in the market, empirical observations such as Bouchaud et al. (2018) find that only a few orders are bigger than the available volume at the best quote. This indicates that traders try to adapt their order volumes to the available quotes.

\vspace{5mm}
\subsection{The Dataset}

\vspace{5mm}
\subsubsection{Lobster Data}

\vspace{5mm}
\noindent
In our research, we will be specifically analysing trades and quotes data from the NASDAQ electronic stock exchange (for more information on NASDAQ please see Appendix C.1). The time-series data has been provisioned from the academic database LOBSTER (Limit Order Book System Efficient Reconstructor) that offers on-demand LOB data, reconstructed from NASDAQ’s Historical TotalView-ITCH files. TotalView-ITCH is a standard NASDAQ data feed which illustrates the full depth of the order book (resting limit orders), as well as all market events. This data feed is consumed in a form of message files which record every state change to the order book, as opposed to recording timely snapshots. TotalView-ITCH message data contains all visible order activities, such as submission, cancellation, and matching of limit orders. Therefore, as outlined in the paper by LOBSTER development team, the platform can reconstruct LOB for a NASDAQ stock at any required book depth level for the specified period (Huang, Lehalle and Rosenbaum, 2015). For a detailed procedure on how LOBSTER reconstructs LOB data, please refer to Appendix C.2.

\vspace{5mm}
\subsubsection{Output Format}
\vspace{5mm}
\noindent
Market activity for a stock on a trading day is organised into two LOBSTER output files: 

\begin{enumerate}

    \item ‘Message’ file contains every arriving market and limit orders as well as cancellations and updates – in other words, all \textit{trades} data 
  
    \item ‘Orderbook' file depicts the evolving state of the LOB. It describes how the total volume of buy or sell orders at corresponding price level changes after each market event. This information is displayed as a list of all successive \textit{quotes}

\end{enumerate}

\noindent
For every entry in the message file, there is a corresponding record in the orderbook file that describes how the order book advanced immediately after the message file event. 

\vspace{5mm}
\begin{table}[hbt!]
\centering
\begin{tabular}{c c c c c c rc}
\hline\hline 
& Time–stamp    & Event Type        & Order ID            & Size          & Order Price  & Direction \\ \midrule
 & 43955.2422      & 4                 & 140339446         & 5              & 2158800          & 1           \\

 & \textbf{ 43955.2426}      & \textbf{4}           & \textbf{ 140339446}     & \textbf{10}        & \textbf{ 2158800}         & \textbf{1}   \\
 & \textbf{ 43955.2426}      & \textbf{4}          & \textbf{ 140339446}    & \textbf{75}         & \textbf{ 2158800}          & \textbf{1}    \\

 & 43955.2426      & 3                 & 140339455         & 100            & 2159600          & -1             \\
 & 43955.2426      & 1                 & 140339468         & 100            & 2159500          & -1              \\
 & 43955.2442      & 3                 & 140339468         & 100            & 2159500          & -1               \\
 & 43955.2468      & 1                 & 140339505         & 100            & 2158900          & -1                \\
 & 43955.2484      & 5                 &     0             & 300            & 2158800          & -1                 \\
 & 43955.2512      & 3                 & 140339505         & 100            & 2158800          & -1                  \\
 & 43955.2513      & 1                 & 140339541         & 100            & 2159600          & -1                   \\

\hline\hline 
\end{tabular}
\caption{Sample entries in ‘message’ file}
\label{table:sampleMsg}
\end{table}


\vspace{5mm}
\begin{table}[hbt!]
\centering
\begin{tabular}{c c c c c}
\hline\hline 
 & Ask Price    & Ask Volume          & Bid Price          & Bid Volume   \\ \midrule
 & 2159600         & 100               & 2158800           & 85      \\
 
 & 2159600         & 100               & \textbf{ 2158800}           & \textbf{75}      \\
 & 2159600         & 100               & \textbf{ 2158300}           & \textbf{20}      \\
 
 & 2160800         & 100               & 2158300           & 20       \\
 & 2159500         & 100               & 2158300           & 20        \\
 & 2160800         & 100               & 2158300           & 20         \\
 & 2158900         & 100               & 2158300           & 20          \\
 & 2158900         & 100               & 2158300           & 20           \\
 & 2160800         & 100               & 2158300           & 20            \\

\hline\hline 
\end{tabular}
\caption{Sample entries in the ‘orderbook’ file at Level 1}
\label{table:sampleEntries}
\end{table}


\vspace{5mm}
\noindent
Tables 2 and 3 correspond to snapshots of message and orderbook files for the same stock during the same time (measured in the number of market events). An event of type 4 at the timestamp of 43955.2426 (Table 2) illustrates the execution of a buy LO at the ask price 2158800. The size of this order is 10 shares; thus, we observe corresponding Bid Volume entry in the orderbook decreasing by 10. Furthermore, the next arriving MO (represented by LO execution at 43955.2426 in Table 2) consumes all available liquidity of 75 shares and the best bid price changes permanently to 2158300 (next best price level in the LOB). In this work, we implement an algorithm that detects this kind of price changing events and explores the causal relationship between market impact and observed LOB features.

\vspace{5mm}
\noindent
In summary, the information contained in output LOBSTER files has the below properties:

\begin{itemize}
  \item All events have timestamps of seconds after midnight with the precision of at least milliseconds (nanoseconds depending on the requested period).  When a market order is matched against several limit orders, each matching is recorded separately (both in message and orderbook files) but with the same timestamp. This allows reconstruction of the initial market order volume.
  
  \item There are seven types of market events that are recorded in LOBSTER data (see Table 4 below). For this study, we are interested in market orders which correspond to event types 4 and 5 – execution of either visible or hidden limit orders. A more detailed explanation of why other types of events are out of scope follows in the Empirical Findings chapter.
  
  \item Order ID corresponds to the unique order reference number. Zero reference number corresponds to a hidden limit order. Note that order ID is distinct from an ID of a trader/ broker who submitted the order. In other words, knowing the order ID does not allow to reconstruct the ownership of trades, but provides information about a lifespan of a single trade.
  
  \item Size of a trade is measured in the number of shares.
  
  \item Price is depicted in dollars times 10000. For instance, 2158800 corresponds to \$215.88. 
  
  \item Direction indicates whether a buy or a sell order has been executed:
  
  \begin{itemize}
     \item  \textbf{-1}: Execution of sell LO, therefore, a MO to buy has been matched at ask price
     \item  \textbf{ 1}: Execution of buy LO, therefore, a MO to sell has been matched at the bid price 
   \end{itemize}
   
   The prevailing majority of microstructure literature that we have consulted during our study adopts the opposite nomenclature. To remain consistent with the subject expertise we follow earlier examples and use trade sign of -1 to indicate sell MO and 1 for buy MO.
  
  \item Orderbook file entries are composed of the ask price and the corresponding volume, as well as the bid price and its volume. The best prices are Level 1 offerings – the cheapest price to buy at (ask), and the biggest price to sell at (bid) from a viewpoint of a trader submitting a MO. Typically, orderbook has several price levels: from the best Level 1 to the second-best, the third best and so on. However, as we recall from the earlier discussion of the LOB, in most electronic exchanges today, an order is re-routed to other markets if the available volume at the best price is less than the order size. Hence, nowadays, an order rarely “walks the book” (being matched with LOs at a worse price on a different price level). Following suit, we will be looking primarily on the Level 1 price changes in the LOB. 
  
\end{itemize}

\vspace{5mm}
\begin{table}[hbt!]
\centering
\begin{tabular}{c c   lc  }
\hline\hline 

 & Event Number    & Event Type                                                         \\ \midrule
 & 1                 &  Submission of a new limit order                                   \\
 & 2                 & Cancellation (partial deletion of a limit order)                   \\
 & 3                 & Deletion (total deletion of a limit order)                          \\
 & 4                 & Execution of a visible limit order                                   \\
 & 5                 & Execution of a hidden limit order                                     \\
 & 6                 & Cross Trade (Auction Trade)                                            \\
 & 7                 & Trading halt indicator                                                  \\

\hline\hline 
\end{tabular}
\caption{Event Types in LOBSTER data}
\label{table:eventTypes}
\end{table}

\subsubsection{Observed Stocks}
\vspace{5mm}
\noindent
The analysis is conducted on four NASDAQ stocks – large tick SIRI and EBAY; and small tick TSLA and PCLN. 
Conventionally, the tick size is the smallest movement in the price of an asset. On NASDAQ each asset is traded in its own book with the tick size of \$0.01. Size of the tick is uniform across all NASDAQ listed securities, despite their prices varying significantly across several magnitudes. \textit{Relative tick size}, then, is defined as a ratio between dollar tick size and the price of a stock. Securities with smaller traded prices have a large relative tick size, while those that trade at higher prices per share have a smaller ratio between tick size and stock price. Large tick stocks are known to have the bid-ask spread almost always equal to one tick, whilst smaller tick securities usually have spreads of a few ticks (Eisler, Bouchaud and Kockelkoren, 2012). The relative tick size and spread indicate how actively an asset is traded on an exchange.  

\vspace{5mm}
\noindent
We have chosen to work with different (in this regard) stocks to be able to quantitatively observe the influence of a tick size on trading features. 

\vspace{5mm}
\noindent
The initial objective of our study was to use trades and quotes data for the selected stocks during the first six months of 2015 ($2^{nd}$ of January to $30^{th}$ of June 2015). Depending on the stock, the average number of daily market events varies from 8,000 to over 200,000. Among these, we are interested in the price impact of MOs specifically. Table 5 lists an average daily number of MOs for each asset.

\vspace{5mm}
\vspace{5mm}
\begin{table}[hbt!]
\centering
\begin{tabular}{c lc c}
\hline\hline 

 & Ticker               & Number of MOs  \\ \midrule
 & \textbf{SIRI}                 &  624            \\
 & \textbf{EBAY}                 &  3540            \\
 & \textbf{TSLA}                &  3924             \\
 & \textbf{PCLN}                &  1333              \\

\hline\hline 
\end{tabular}
\caption{Average daily number of MOs for each stock}
\label{table:aveDailyMO}
\end{table}


\vspace{5mm}
\noindent
During the first six months of 2015, there have been 124 NASDAQ trading days which gives us a total number of relevant MO events of the order of ${10}^5$. This implies that our dataset is sufficiently large, ensuring significant robustness in any statistical findings. In the later empirical discussion, we elaborate further on the nature of what we have learned from the data and how our results compare to other studies that use similar LOBSTER dataset.

\vspace{5mm}
\subsection{Method Development}

\vspace{5mm}
\noindent
The methods employed in this research are solely based on the principles of the scientific method. The process of the scientific method begins with the formulation of a question based on observations. This allows a hypothesis to be formed that may explain an observed phenomenon. In this instance, a hypothesis may be \textit{"do large orders impact prices?"} The null hypothesis is large orders of size (traded volume) $Q > \mathbb{Z}$ do not impact prices, where $Q$ denotes the order size.

\vspace{5mm}
\noindent
After formulation of a hypothesis, it is up to the scientist to disprove the null hypothesis and demonstrate that orders of a predefined defined (large) size do in fact impact prices. To carry this out a prediction must be defined. To prove or disprove the hypothesis, the prediction is subject to testing. 

\vspace{5mm}
\noindent
The results of the testing procedure will provide a statistical answer upon whether the null hypothesis can be rejected at a certain level of confidence. If the null hypothesis cannot be rejected, which implies that there was no discernible relationship between the size of an order and the price impact, it is still possible that the hypothesis is (partially) true. A larger set of data, a transformation of variables or incorporation of additional information can be used to optimise and improve the level of significance. This is the process of analysis and it seeks to reject the null hypothesis after refinement. For more details on \textit{statistical inference} please refer to Appendix C.5 and Appendix C.6.

\vspace{5mm}
\subsubsection{Machine Learning}

\vspace{5mm}
\noindent
Machine learning (ML) is a subfield of Artificial Intelligence (AI) that facilitates automated methods of data analysis. More broadly, ML defines a series of adaptive computational algorithms that automatically detect patterns in \textit{multi-dimensional datasets} (panel or time-series data with a large number of observed variables) and makes use of these uncovered patterns to predict values of interest or perform other kinds of decision making under uncertainty (known as generalisation). 

\vspace{5mm}
\noindent
The conventional tool of statistical analyses in finance – econometrics – mainly focuses on multivariate linear regression. Linear regression does not have memory (standard econometric models do not learn), thus fails to improve its performance with new observations. Consequently, econometric regression analysis often falls short in understanding the full spectrum of informational content present in the data. Moreover, when applied to complex problems, more traditional statistical tools often suffer from various limitations (i.e., the “curse of dimensionality” – the model becomes mathematically intractable when dealing with a large number of explanatory variables/ features), whilst ML algorithm can learn patterns in high-dimensional space without being specifically directed. In this regard, ML methods do not replace the theory and conventional wisdom of econometrics but simply enhance and guide them.

\vspace{5mm}
\noindent
The defining attribute that differentiates ML from such conventional tools is the algorithms’ ability to \textit{learn} (i.e., improve predictive accuracy with time) from experience with respect to a task and performance measure (Mitchell, 1997). Here, \textit{experience} refers to the past information available to a learner (typically taking the form of electronic training data). Thus, ML algorithms learn patterns in a high-dimensional space without being explicitly directed (i.e., programmed with pre-specifications). Such algorithms are incredibly diverse, ranging from more traditional statistical models that emphasise inference to \textit{deep neural network} architecture that excel at highly complex classification and predictive tasks. As the success of learning is greatly dependant on the data used, ML most closely relates to statistics and data mining, though it has a different emphasis and vocabulary.  

\vspace{5mm}
\noindent
ML tasks are generally classified into three broad areas: \textit{Supervised Learning}, \textit{Unsupervised Learning} and \textit{Reinforcement Learning}. In the current study, we specifically focus on Supervised Learning, which is the most commonly used class of ML. Please refer to Appendix C.4 for information on other forms of ML.

\vspace{5mm}
\noindent
Supervised Learning describes a set of predictive learning models for which the presence of outcome variables guides the learning process. That is, data annotated with values such as categories (as in supervised classification) or numeric responses (as in supervised regression) facilitates external supervision of the learning process. The objective is to learn a mapping from inputs $x$ to outputs $y$, given the \textit{labelled} set of input-output pairs $\mathcal{D}= {(x_i,y_i)}_{i=1}^N $. Here $\mathcal{D}$ represents the training set, and $N$ is the number of training examples. Given the set of labelled examples, the algorithm is \textit{trained} on the data and learns which predictive factors are most influential to the responses. When applied to new unseen datasets, supervised learning algorithms attempt to make predictions based on their prior training experience. In the simplest setting, each training input $x_i$ corresponds to a D-\textit{dimensional} vector of values that correspond to various attributes of the observed phenomenon. These are known as \textit{features} and could range from a time-series of observations to more complex structured objects such as an image or a molecular shape.

\vspace{5mm}
\noindent
Similarly, the form of the output dependent variable (known as the response variable) can in principle take on any structure, however, most methods assume that $y_i$ is a categorical variable from some finite set $y_i\ \in{1,\ldots,\ C}$ or $Y_i\ \in\ \mathbb{R}$ (a real-valued continuous scalar). 
This research makes explicit use of regression-based supervised algorithms, for which we will delineate the basic mechanisms below.

\clearpage
\vspace{5mm}
\noindent
\textbf{Multi-Linear Regression}

\vspace{5mm}
\noindent
Linear regression is a familiar and widely adopted statistical technique. This asserts that a continuous scalar response $y$ is the result of a linear combination of its feature inputs $x$. Unlike multivariate linear regression, where the model predicts multiple correlated dependent variables given multiple input variables, \textit{multi-linear} models output a single real value scalar. That is:

\begin{eqnarray}
y\left(x\right)=\ \beta^T x +\ \epsilon=\ \sum_{j=1}^{D}{\beta_j x_j +}\epsilon\
\end{eqnarray}

\vspace{5mm}
\noindent
Where $\beta^T,\ x\in\mathbb{R}^{p+1}\ $ and $\epsilon\ \sim \mathcal{N}(\mu,\ \sigma^2)$. That is, $\beta^T$ and $x$ are both real-valued vectors of dimension $p+1$, representing the inner scalar product between the input vector ${x}$ and the models' weight vector $\beta^T$.$\epsilon$, the residual error between our linear predictions and the true response, is normally distributed with mean $\mu$ and variance $\sigma^2$. The multi-linear regression model makes the strong assumption of \textit{independent and identical distribution} (IID) of errors. That is, the distribution of the errors is identical across observations and the errors are independent of each other (knowing one error does not assist in forecasting the next error). When plotted, such a distribution produces the well know Gaussian (normal) distribution. To make the connection between linear regression and Gaussian distributions more explicit, we can express the model in the form: 

\begin{eqnarray}
p\left(y\middle|x,\theta\right)=\ \mathcal{N}(\mu\left(x\right),\ \sigma^2\left(x\right))\ 
\end{eqnarray}

\vspace{5mm}
\noindent
This makes it clear that the model is a conditional probability density. In the simplest case, we assume $\mu$ is a linear function of $x$, thus $\mu=\ \beta^T x$; and the noise (as measured by variance) is fixed as $\sigma^2(x)=\ \sigma^2$. In this case, $\theta=(\beta,\ \sigma^2$ are the parameters of the model.

\vspace{5mm}
\noindent
Multilinear regression forms the basis of many of the more sophisticated supervised ML techniques employed. From the defined model specification, it is intuitive to ask how the coefficient $\beta$ is estimated. Further, we introduce two methods for optimal $\beta$ estimation.

\vspace{5mm}
\vspace{5mm}
\noindent
\textbf{Least Squares}

\vspace{5mm}
\noindent
The \textit{Least Squares} is the most common approach to learn the parameters of a hyperplane by approximating coefficients that \textit{best fit} the generated output for a specified input set. The difference between the model’s prediction and the actual outcome for a given data point denotes \textit{the residual}, whereas the deviation of the model from the true population output is called \textit{the error}. The method of \textit{Ordinary Least Squares} (OLS – a type of \textit{Least Squares} used for estimating the parameters in a \textit{linear} regression model) entails taking each vertical distance from a data point to the regression line (residual), squaring this distance (taking an absolute value by disregarding the sign) and then minimising the total sum of squared residuals, see Figure 2.
In more formal terms, the least square estimation method chooses the coefficient vector $\beta$ to minimise the residual sum of squares (RSS, also known as the sum of squared errors SSE) so that the model fits the data as closely as possible. Hence, the least-squares coefficients $\beta^{LS}$ are computed as 

\begin{eqnarray}
\underset{\beta^{LS}}{\arg\min} = RSS(\beta)\ {:= } \sum_{i=1}^{N}(y_i\ -\ \beta^T x_i)^2
\end{eqnarray}

\vspace{5mm}
\noindent
The vertical distances are usually minimised as opposed to the horizontal distances or those taken perpendicular to the line. This arises as a result of the assumption that $x$ is fixed in repeated samples so that the problem becomes one of determining the appropriate model for $y$ given (or conditional upon) the observed values of $x$. 

\vspace{5mm}
\vspace{5mm}
\begin{figure}[hbt!]
    \centering
    \includegraphics[height = 7cm, width=9cm]{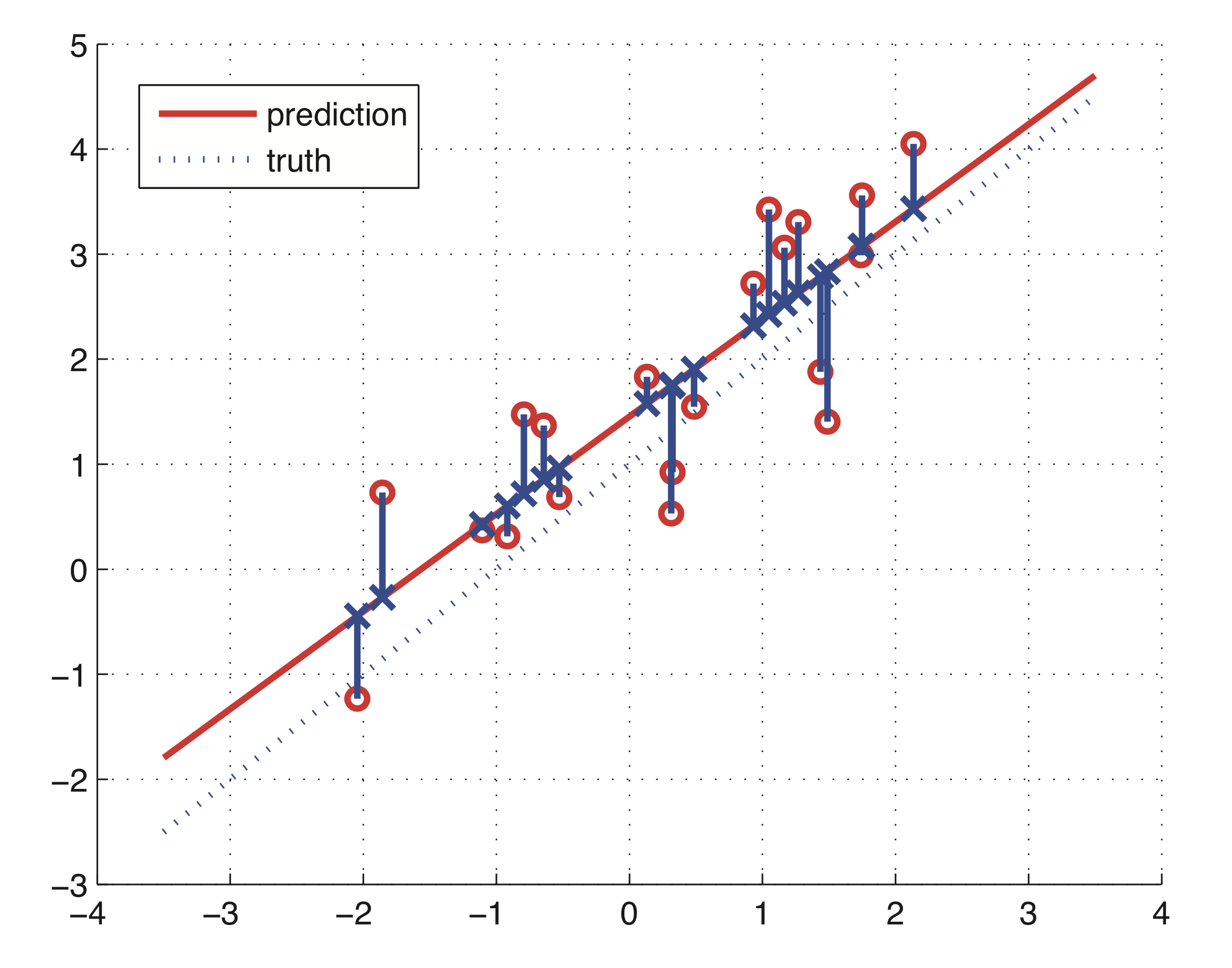}
    \caption{Method of OLS fitting a line (described by the model) to the data by minimising the sum of squared residuals (Murphy, 2012).}
    \label{fig:olsFit}
\end{figure}

\vspace{5mm}
\noindent
For simplicity, the latter term can be expressed in matrix form. By defining the \break $N\times(p+1)$ matrix $x$, it is possible to write the RSS term as:

\begin{eqnarray}
RSS(\beta)\ =\ {(y\ -\ x\beta)}^T(y -\ x\beta)
\end{eqnarray}

\noindent
This term is now differentiated with respect to (w.r.t.) the parameter variable $\beta$:

\begin{eqnarray}
\frac{\partial RSS}{\partial\beta}=\ -2 x ^T(y\ -\ x\beta)
\end{eqnarray}

\vspace{5mm}
\noindent
A key assumption about the data is made here: the matrix $x^T x$ must be positive-definite, which is only true if there are more observational datapoints than there are dimensions. If this does not hold (as is often the case in high-dimensional data settings) then it is not possible to find a unique $\beta$ coefficient, thus the above matrix equation cannot hold. Under the assumption of a positive-definite $x^T x$ the partial differential (PDE) is set to zero and solved for  $\beta$:

\begin{eqnarray}
x^T\left(y- x \beta\right)=0
\end{eqnarray}

\vspace{5mm}
\noindent
The solution to this matrix equation provides ${\hat{\beta}}_{OLS}$

\begin{eqnarray}
{\hat{\beta}}_{OLS}\ =\ {(x^T x)}^{-1} x^T y
\end{eqnarray}

\vspace{5mm}
\noindent
\textbf{Maximum Likelihood}

\vspace{5mm}
\noindent
An alternative to OLS, the \textit{Maximum Likelihood Estimator} (MLE) is an optimisation process to estimate the coefficients of a statistical model given a particular batch of data by maximizing the \textit{likelihood function} (defines how \textit{likely} a set of observations is to occur given the model parameters). The estimator differs from probabilities in that it is not normalised to range from 0 to 1. MLE is a directed algorithmic search through a high-dimensional set of possible parameter choices, attempting to answer the question: \textit{If the data were to have been generated by the model, what parameter choices were most likely to have been used}? (Murphy, 2012).

\noindent
This reduces to a conditional likelihood problem of seeing the dataset $\mathcal{D}$, given a specific set of parameters $\theta$. The value sought is the $\theta$ that \textit{maximises} $p(\mathcal{D}|\theta)$. This can be framed as searching for the mode of the $p(\mathcal{D}|\theta)$ denoted by $\hat{\theta}$,  expressed as 

\begin{eqnarray}
\hat{\theta} \triangleq \underset \theta{\arg\max} \log p(\mathcal{D}|\theta)
\end{eqnarray}

\vspace{5mm}
\noindent
In linear regression problems, we often assume that the errors are IID. This simplifies the solution of the log-likelihood by making use of properties of the natural logarithms, permitting us to express it as

\begin{eqnarray}
\ell\hat{\theta} \triangleq \log{p(\mathcal{D}|\theta)}=\ \sum_{i=1}^{N}\log{p(y_i|x_{i,\ }\ \theta)}
\end{eqnarray}

\vspace{5mm}
\noindent
In the case of normal distribution of residuals, maximising the log-likelihood function returns the same parameter solution as Least Squares.

\vspace{5mm}
\subsubsection{Goodness Of Fit}

\vspace{5mm}
\noindent
It is often desirable to measure how well the regression line (described by the model) fits the data (i.e., explains the variation in the outcome response function). \textit{Goodness-of-fit} statistical measures rigorously assess the quality of the \textit{sample regression function} (SRF) specifications, permitting us to select model designs that best estimate the true relationship between observed variables. Prominent goodness-of-fit measures, including the coefficient of determination (\textit{R-squared} ($R^2$)) and \textit{adjusted R-squared}, are based on maximising the outcome of least-square estimates.

\begin{itemize}
  \item 	$R^2$ measures the proportion of total variation in the observed data explained by the model and ranges from 0 to 1. $R^2$ is computed as  $R^2 =  1-\frac{RSS}{TSS}$ , where TSS is the total sum of squared deviations of the outcome from its mean, given by
  
\begin{eqnarray}
  TSS =\sum_{i} {(y_i -\frac{1}{N} \sum_{i=1}^{N}y_i)^2}
\end{eqnarray}

\vspace{5mm}
and RSS is the \textit{residual sum of squares} (the sum of all of the squared differences between the actual datapoints and the corresponding estimated values), expressed as 

\begin{eqnarray}
  RSS\ =\sum_{i}{{(y}_i-\ {\hat{y}}_i)}^2
\end{eqnarray}

\vspace{5mm}
The objective is to maximise $R^2$ - the closer the value of $R^2$ is to 1 the better the regression model fits the data (when compared to the mean of the observation). $R^2$ only illustrates how one variable explains the observation. When more explanatory variables are added to the model an alternative adjusted $R^2$ should be used.

    \item \textit{Adjusted} $R^2$ penalises for increasing the feature space with non-significant explanatory variables, by reducing RSS to produce a superior goodness-of-fit. 

\end{itemize}

\vspace{5mm}
\noindent
Both $R^2$ and adjusted $R^2$ are not ideal for time-series predictive modelling as they only reflect the models’ precision in fitting past values, which is not indicative of future predictions.

\vspace{5mm}
\noindent
Alternative prominent goodness-of-fit measures are discussed in Appendix C.7.

\vspace{5mm}
\subsubsection{Feature Engineering}

\vspace{5mm}
\noindent
ML algorithms are only capable of learning if the training set contains sufficient relevant variables in the feature space and minimum irrelevant ones. A critical aspect of a success ML initiative is producing a good set of features for which the algorithm can be trained on. This process is known as  \textit{feature engineering} and involves: 

\begin{enumerate}

    \item 	\textit{Feature Extraction} – combining existing features to produce a more useful variable
  
    \item 	\textit{Feature Selection} – selecting the most useful/ relevant features to train the algorithm on amongst an existing set of features.

\end{enumerate}

\subsubsection{Cross–Validation}

\vspace{5mm}
\noindent
It could be stated that the defining quality of a successful ML algorithm is its ability to generalise across future unseen datasets. To ensure accuracy and robustness of a model it is necessary to define a loss function which measures the predictive precision of our model, as a function of the \textit{generalisation error}. In a regression setting, a common loss function is given by the \textit{Mean Squared Error} (MSE) - a smaller MSE means the estimate is more accurate. It is defined as:

\begin{eqnarray}
  MSE \ {:= } \frac{1}{N} \sum_{i=1}^{N}{|{y}_i - {\hat{y}_i|}^2 }
\end{eqnarray}

\vspace{5mm}
\noindent
This states that the generalisation error of a model, given a particular set of data, is the average of squared differences between the training values $y_i$ and their associated estimates $\hat{y}_i$. This function heavily penalises estimate values that differ greatly from the true observations (by squaring the differences). A small value of a loss function signifies that the errors are not substantial and, thus, the model, theoretically, will  perform similarly when exposed to unseen data.

\vspace{5mm}
\noindent
It is important to note that the \textit{standard MSE} value is computed only on the training dataset (the data for which the model was fitted on) and is, therefore, referenced as the \textit{training MSE}. This value is of little practical importance as it is merely representative of the past predictive accuracy; we are more concerned about how well the model performs given values of new unseen data. This is known as generalisation performance. Given a new prediction value   $x_0$ and a true response $y_0$, we look to take the expectation across all such new prediction values, giving us the \textit{test MSE}:

\begin{eqnarray}
  MSE_\textit{test} \ {:= } \mathbb{E} \left[{(y}_0 - {\hat{f}(x_0))}^2\right]
\end{eqnarray}

\vspace{5mm}
\noindent
Where the expectation is taken across all unseen predictive pairs ${(y}_0,x_0)$. The objective is to select the model that yields the lowest MSE. To do this, we can increase the \textit{flexibility} of the model (the degrees of freedom available to the model to \textit{fit} to the training dataset). In this regard, a linear model is very inflexible (it only has 2 degrees of freedom), whereas a polynomial is highly flexible (it can have many degrees of freedom).

\vspace{5mm}
\noindent
However, if the loss function is minimised too severely, then generalisation performance of the model can decrease substantially. This is a major known concern and is referred to as the \textit{bias-variance trade-off} – indicating that the model has been \textit{over-fit} to the training data, and has not \textit{learned} the \textit{general form} of the response function. Whilst increasing the degrees of freedom helps the model adapt to more complex datasets, it is extremely important to recognise that there is an increased likelihood of \textit{over-fitting}, and a clear warning that the model is more closely aligned to minor variations in the training input set (\textit{noise}) than any underlying true signal.

\vspace{5mm}
\noindent
A technique to mitigate the effects of the bias-variance trade-off is to use a separate validation dataset (different from training data) by randomly dividing observations in two partitions - \textit{in-sample} (used for model training) and \textit{out-of-sample} (used to examine the predictive estimates against the true values) sets. However, if we don’t have a separate set of reference out-of-sample data, we can adopt a technique known as \textit{cross-validation} (CV). In the current study we make use of a specific implementation of CV known as \textit{k-Fold cross validation}. \textit{k}-Fold seeks to divide $n$ observations of the data into $k$ mutually exclusive and approximately equal sized \textit{folds} (subsets). This is repeated $k$ times, with each iteration \textit{holding out} a fold as a validation set, whilst the remaining $k-1$ are left for training. This permits us to calculate an overall estimate ${\rm CV}_k$, giving us the average of all individuals mean squared errors, ${\rm MSE}_i$. More formally, ${\rm CV}_k$ is defined as follows:

\begin{eqnarray}
  {\rm CV}_k = \frac{1}{N} \sum_{i=1}^{N}{\rm MSE}_i
\end{eqnarray}

\clearpage
\noindent
Due to computational expenses, the optimal value for $k$ has been empirically proven to be $k\ =5$ or $k\ =10$.

\begin{figure}[hbt!]
    \centering
    \includegraphics[height = 7cm, width=9cm]{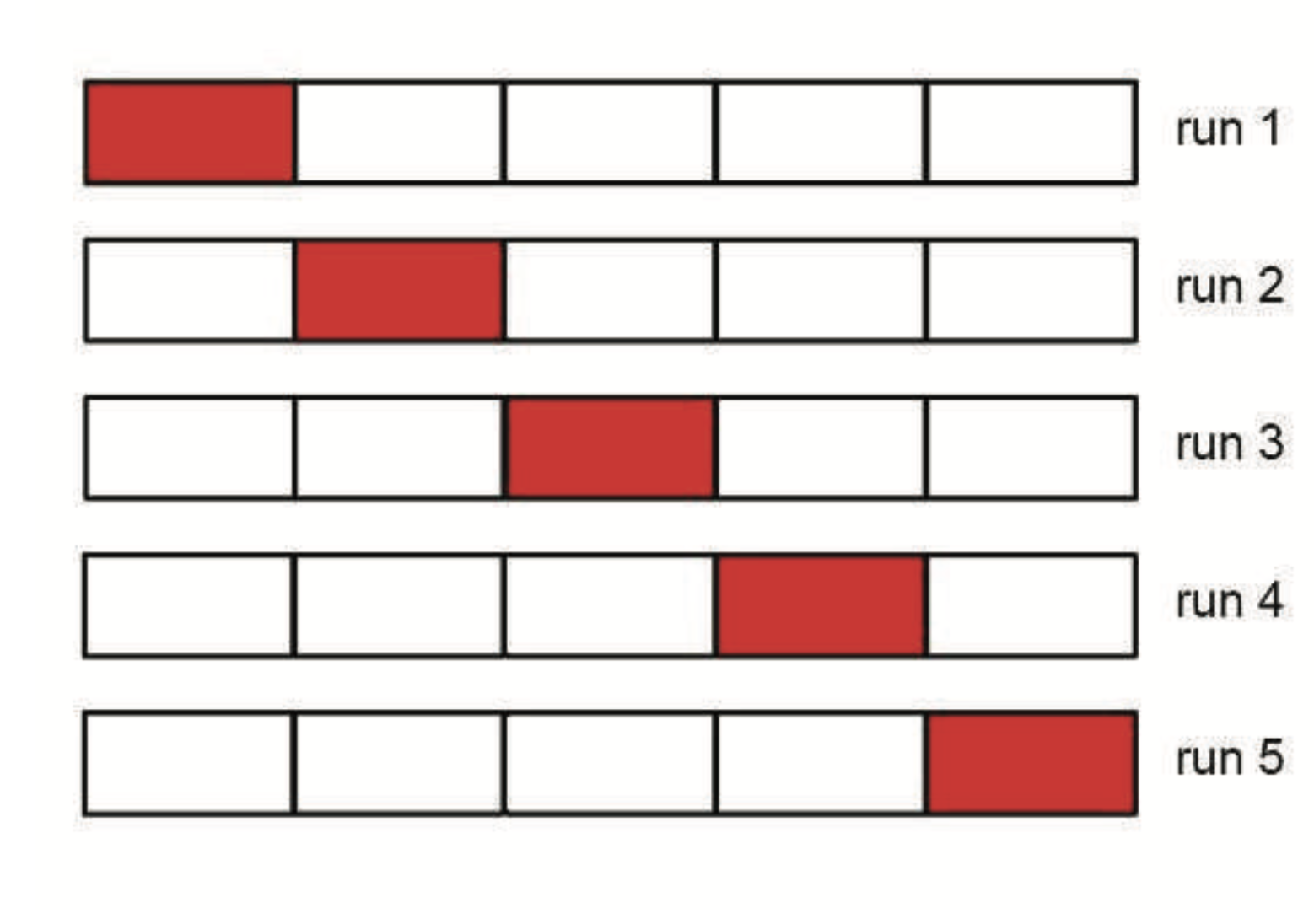}
    \caption{Schematic of 5-fold cross validation (Murphy, 2012).}
    \label{fig:kFold}
\end{figure}

\section{Empirical Study}
\vspace{5mm}
In this section, we define the methodology for the price impact analysis, while simultaneously introducing several key properties that are relevant to our investigation. In the earlier background and literature review section, we have discussed several popular market impact models such as the linear Kyle model and Square Root empirical functional form. Here, we develop a further understanding of price dynamics by exploring the functional relationships between change in price (impact) and state of the world before and after.  

\vspace{5mm}
\noindent
Please note that the data used for our statistical analysis has been pre-processed to remove outliers. We further reconstruct original MOs that are represented as several LO executions within the LOBSTER dataset. Appendix C.3 describes in detail the applied pre-processing procedure.

\vspace{5mm}
\subsection{Modelling Market Impact}
\vspace{5mm}
\noindent
In broad terms, price impact can be defined as an incremental change in price caused by the execution of market buy orders, or a similar drop in price caused by the sell orders. Statistically, market impact relates to the positive correlation between the incoming MO sign and the price change that follows immediately, or sometime after MO execution (Bouchaud et al., 2018). 

\vspace{5mm}
\noindent
Therefore, in our study of price impact, we focus on the consequences of MOs. We recognise that all kinds of market events may impact prices, but we choose to concentrate on the effect of MOs because it is conceptually and operationally intuitive to study initially – i.e., it suffices to use trades and quotes data from the exchange.

\vspace{5mm}
\noindent
In combination with earlier remarks about submitted order sizes rarely exceeding available volume at the opposite side best quote (LOBSTER output discussion in Section 3.2.2), we postulate the following two assumptions of our empirical analysis:

\begin{enumerate}
  \item Market orders never exceed available liquidity in the market
  \item Impact of arrivals and cancellations of LOs can be neglected
\end{enumerate}

\vspace{5mm}
\noindent
These assumptions will help us understand the high-level dynamics of interactions in the marketplace without introducing unnecessary complexity to the decision-making process.  

\vspace{5mm}
\noindent
With this in mind, we can begin to measure the impact of trades on prices.

\vspace{5mm}
\subsubsection{Unconditional Lag-1 Impact}
\vspace{5mm}
\noindent
Bouchaud et al. (2003) in their paper \textit{“Fluctuations and response in financial markets: the subtle nature of `random' price changes”} postulate that the simplest quantity, that can assist in the study of price changes, is the mean squared fluctuation of the prices between the given trade and the execution of the next one (correspondent to the execution of MOs in the context of LOB trading). They functionally define this degree of fluctuation $D\left(l\right)$ as:

\begin{eqnarray}
D\left(l\right)\ {:= } \langle(m_{t+1}-m_t)\rangle
\end{eqnarray}

\vspace{5mm}
\noindent
Where $m_t$ is the mid-price immediately before the $t^{th}$ MO:

\begin{eqnarray}
m\left(t\right)\ {:= } \frac{1}{2} (a\left(t\right)+ b\left(t\right))
\end{eqnarray}

\vspace{5mm}
\noindent
$a\left(t\right)$ and $b\left(t\right)$ are the corresponding ask and bid prices.

\vspace{5mm}
\noindent
Whilst $D\left(l\right)$ measures the degree of diffusive behaviour of market prices, authors propose a better alternative to examine specifically the effect of trading on the change in prices - \textit{lagged response function} $\mathcal{R}()$. $\mathcal{R}(1)$  is the lagged by 1 response function that measures the average difference between the mid-price just before the arrival of an original MO and the mid-price just before the arrival of the next MO:

\begin{eqnarray}
\mathcal{R}(1) {:= } \langle \varepsilon_t \cdot(m_{t+1} - m_t)\rangle_t,
\end{eqnarray}

\vspace{5mm}
\noindent
In contrast to the previous diffusion estimation, the response function includes the order sign $\varepsilon_t$  to account for the direction of MO. For instance, if a market sell order (direction -1) causes an asset price to decrease, the change in mid-price will be negative. To compensate for this, we incorporate the order sign -1 (given by $\varepsilon_t$). This allows the correct estimation of correlation between MO and subsequent price change. Specifically, $\mathcal{R}\left(1\right)$ measures how much, on average, the price increases given a buy order (or how sell order moves the price down).  

\vspace{5mm}
\noindent
The lag-1 response is calculated as the empirical average of $\mathcal{R}\left(1\right)$ over all consecutive MOs. This definition can be further extended to look at the relationship of MOs beyond lag-1; or conditioned on extra variables. However, initially, we explore the statistical properties of $\mathcal{R}\left(1\right)$ as defined above.

\vspace{5mm}
\clearpage
\noindent
\textbf{Order Splitting}

\vspace{5mm}
\noindent
Lag-1 response function measures market reaction to a single MO. In reality, traders wishing to execute large quantities have to split their orders into child MO to reduce the impact by matching available liquidity at the corresponding price level (please see Appendix B.1 for more detail). To study the dynamics of such meta-orders, it is necessary to know which child orders belong to the same meta-order. Such levels of information can only be found in specialised and/ or proprietary data sets. Due to such restrictions, this paper solely focuses on the impact of single independent MOs.

\clearpage
\subsubsection{Engineering Response Function}
\noindent
We calculate the lag-1 response function for each of our four stocks during the first 6 months of 2015 using \textit{uncond\_market\_impact.py} script provided in the Appendix E. The general idea is demonstrated in algorithm 1 below:

\vspace{5mm}
\begin{algorithm}
\small
\DontPrintSemicolon 
\KwIn{A DataFrame set of trades and quotes entries $D = \{d_1, d_2, \ldots, d_r\}$, where $\forall t: d_t$ entry contains the following information about the market event: $\{ Time, Event Type Code, Order ID, Volume, Price, \newline Direction, Ask Price, AskVolume, Bid Price, BidVolume, \newline Mid-price, Spread \}$}

\KwOut{$\langle s \rangle, R(1), \Sigma_r, N_{MO}$ }
$\langle s \rangle, R(1),\Sigma_r, N_{MO} \gets 0$\;
$m_t, m_{t+1}, V(1)\gets 0$\;
\For{$i = 0, \ldots,$ \textbf{length} $(D)$}{
  \uIf {$d_i[EventType] = 4$ \textbf{or} $d_i[EventType]=5$}{
  $N_{MO} += 1$\;
  $\langle s \rangle += d_{i-1}[Ask]-d_{i-1[}Bid]$\;
  \uIf{$m_t=0$}{
    $m_t=d_i[Midprice]$\;
  }
  
  $m_{t+1}=d_{i+1}[Midprice]$\;
  $R =$ \textbf{max} $(0, ( m_{t+1}-m_t)*sign)$\;
  $R(1) += R$\;
  $V(1) += ( m_{t+1} - m_t)^2$\;
  
  $m_t=m_{t+1}$\;
  }
}
$\langle s \rangle=\langle s \rangle / N_{MO}$\;
$R(1)=R(1)/ N_{MO}$\;
$V(1)=V(1)/(N_{MO}$\;
$\Sigma_r=\sqrt{V(1)-R(1)^2}$\;
\Return{$\langle s \rangle, R(1),\Sigma_r,N_{MO}$}\;
\caption{Calculate \textit{lag}-1 response}
\label{algo:change}
\end{algorithm}

\vspace{5mm}
\noindent
In addition to $\mathcal{R}\left(1\right)$ the script estimates:

\begin{itemize}
    \item 	$\langle s \rangle$ - average spread just before a MO
    
    \item 	$\sum_\mathcal{R}$ - standard deviation of price fluctuations around the average price impact of a MO
    
    \item 	$N_{MO}$ – number of MOs recorded during the day (excluding the auctions) 
    
\end{itemize}

\noindent
A MO is said to be price-changing if its size is equal to or exceeds the volume at corresponding opposite-side best price in the LOB at the time of MO arrival; that is, MO execution changes the current price of an asset immediately to the next best bid (ask) respectively.  

\vspace{5mm}
\noindent
In summary, the above algorithmic procedure iterates over daily market events and when it finds a MO execution (corresponding to the event of type 4 or 5), it records various statistics. Appendix C.8 explains specifically how the events of type 5 – execution of hidden orders - can affect our measurement of market impact. For each trading days’ set of observations we calculate daily averages, and then use it to estimate descriptive statistics of our data for the entire period.

\clearpage
\subsubsection{Empirical Observations}
\vspace{5mm}
\noindent
The results of our calculations can be found in the below Table 6:

\vspace{5mm}
\begin{table}[hbt!]

\centering
\begin{tabular}{c lc c c c c}
\hline\hline 
 &  Stock  &         $\langle \textbf{s} \rangle$  &$\mathcal{R}\left(1\right)$       &$\sum_\mathcal{R}$      & $N_{MO}$  \\ \midrule
 & \textbf{SIRI}        & 1.09                         & 0.027                        & 0.138                        & 589      \\
 & \textbf{EBAY}        & 1.10                         & 0.384                        & 0.536                        & 3802      \\
 & \textbf{TSLA}        & 10.82                        & 2.05                         & 3.636                        & 3834      \\
 & \textbf{PCLN}        & 75.53                        & 10.09                       & 21.129                       & 1409      \\

\hline\hline 
\end{tabular}

\caption{ $\langle s \rangle$ Average spread; $\mathcal{R}\left(1\right)$ the lag-1 response function for all MOs; $\sum_\mathcal{R}$ standard deviation of price fluctuations around the average price impact of MOs – all expressed in dollar cents; $N_{MO}$ total average number of MO per day between 10:30 and 15:00 on each trading day from $2^{nd}$ of January to $30^{th}$ of June 2015 for four large- and small- tick stocks.}

\label{table:aveSpreadOfR1}
\end{table}


\vspace{5mm}
\noindent
The following can be noted:

\begin{itemize}

    \item 	Lag-1 unconditional response $\mathcal{R}\left(1\right)$ is strongly positive for all stocks. This implies that on average, MOs are followed by a change in price.
    
    \item 	For small tick stocks TSLA and PCLN, $\mathcal{R}\left(1\right)$ seems proportional to the average spread $\langle s \rangle$. This could be explained by our earlier discussion of market makers’ \textit{compensation for adverse selection} – MMs attempt to earn the bid-ask spread but are challenged by adverse price movement caused by the MOs impact. 
    
    \item For large tick stocks, the spread is bound below by (and is often equal to) a single tick as can be noticed for SIRI and EBAY. When the bid-ask spread is this small, neither of the best prices can improve as the two would otherwise merge. If we recall Figure 1 from the earlier discussion of the LOB, there was a gap of 4 ticks between the best ask and bid prices. In that case, submission of a new LO at a price lower than the current ask would update the best quote, thus the mid-price. Large tick stocks, on the other hand, do not have such a gap between the best bid and ask, therefore, the changes in mid-price are not affected by LOs in the same manner. In this regard, Bonart and Gould (2017) argue that large tick stocks are more suited for the analyses of price queue dynamics. 
    
    \item 	For all stocks, the degree of dispersion $\sum_\mathcal{R}$ exceeds the mean impact $\mathcal{R}\left(1\right)$, meaning that $\mathcal{R}\left(1\right)$ varies greatly. This is due to the fact that, as we have observed from the data, MOs can not only change the price in the direction of demand, but have also been found, at times,  to not affect the LOB at all. Moreover, a MO can be followed by the opposite direction change in the mid-price. In this respect, $\mathcal{R}\left(1\right)$ incorporates the effect of the full sequence of market events that have caused the asset price to change including submissions and cancellations of LOs. As pointed out earlier, small tick stocks’ mid-prices are prone to be affected by LOs to a larger degree than large-tick stocks.
    
\end{itemize}

\vspace{5mm}
\subsubsection{2015 Financial Time Series}
\vspace{5mm}
\noindent
Our choice of the observational period was partially influenced by the research of Bouchaud et al. (2018). Authors use LOBSTER data for 120 stocks traded on NASDAQ to measure and predict market impact. They have conducted comparable estimations for our selected securities during 2015.

\vspace{5mm}
\noindent	
When we initially benchmarked our findings for the first six months of 2015 to those of Bouchaud et al. (2018), our response function measurements were somewhat lesser. To understand this phenomenon, we conducted the same experiments on the dataset for the second half of 2015 (to obtain an entire year average as estimated in Bouchaud et al. (2018)).

\vspace{5mm}
\noindent	
The results, presented in Table 7, indicate that trade activity in the second half of 2015 was very distinct from the first six months. This is in line with several public news archives, which report major financial events that took place at the start of June 2015 and followed through to June 2016. During that period, now referred to as \textit{“2015-16 stock market sell-off”}, stock prices around the world declined in value as traders were actively selling for a number of reasons including: slowing growth of Chinese GDP, falling petroleum prices, Greek debt default, sharp rise in bond yields and UK’s decision to leave the European Union among other factors (Randall and Gaffen, 2016).

\vspace{5mm}
\begin{table}[hbt!]
\centering
\begin{tabular}{c lc c c c c}
\hline\hline 
 &  Stock  &         $\langle \textbf{s} \rangle$  &$\mathcal{R}\left(1\right)$       &$\sum_\mathcal{R}$      & $N_{MO}$  \\ \midrule
 & \textbf{SIRI}        & 1.04                         & 0.031                        & 0.156                        & 669      \\
 & \textbf{EBAY}        & 1.10                         & 0.263                        & 0.470                        & 3320      \\
 & \textbf{TSLA}        & 15.05                        & 2.568                        & 4.984                        & 4007      \\
 & \textbf{PCLN}        & 113.17                       & 14.353                       & 32.088                       & 1273      \\

\hline\hline 
\end{tabular}

\caption{ $\langle s \rangle$ Average spread; $\mathcal{R}\left(1\right)$ the lag-1 response function for all MOs; $\sum_\mathcal{R}$ standard deviation of price fluctuations around the average price impact of MOs – all expressed in dollar cents; $N_{MO}$ total average number of MO per day between 10:30 and 15:00 on each trading day from $30^{th}$ of June to $31^{st}$ of December 2015 for four large- and small- tick stocks}

\label{table:aveSpreadOfR1_2}
\end{table}


\vspace{5mm}
\noindent	
The data adheres to the occurrence of such events – the average spread and price dispersion (as measured by the standard deviation) are evidently higher for the period of June to December 2015 (higher than the first half). Although the following year of 2016 is out of the scope of our research, we are confident similar trend continued through.

\vspace{5mm}
\noindent	
When we average the outcome for the two halves of 2015, our results resemble closely those by Bouchaud et al. (2018). Table 8 contrasts our findings to those outlined in \textit{“Trades, Quotes and Prices”} (Bouchaud et al., 2018).

\vspace{5mm}
\begin{table}[hbt!]
\centering
\begin{tabular}{c lc c c c c}
\hline\hline 
&  Stock  &         $\langle \textbf{s} \rangle$  &$\mathcal{R}\left(1\right)$       &$\sum_\mathcal{R}$      & $N_{MO}$  \\ \midrule
 
& \textbf{SIRI}         & 1.06                         & 0.058                       & 0.213                        & 623      \\
&         SIRI*         & 1.06                         & 0.029                       & 0.147                        & 629      \\\\

& \textbf{EBAY}         & 1.10                         & 0.348                       & 0.502                       & 3575     \\ 
&         EBAY*         & 1.10                         & 0.324                       & 0.503                       & 3561      \\\\

& \textbf{TSLA}        & 12.99                        & 2.59                        & 4.49                       & 3932     \\ 
&         TSLA*        & 12.94                        & 2.31                        & 4.31                       & 3921      \\\\

& \textbf{PCLN}         & 94.68                        & 15.30                       & 28.77                       & 1342      \\
&         PCLN*         & 94.35                        & 12.22                       & 26.61                       & 1341      \\

\hline\hline 
\end{tabular}

\caption{ In bold are empirical observations from the work of Bouchaud et al. (2018); in white results of the current study using market data from LOBSTER (measurement units are the same as in Tables 6 and 7)}

\label{table:empiricalComp}
\end{table}


\vspace{5mm}
\noindent	
We prescribe some minor discrepancy in the reported results to the implementation approach. Some further discussion on this is presented in Appendix C.9. 

\subsubsection{Conditioning on Trade Volume}
\vspace{5mm}
\noindent
In our earlier discussion of market impact, we looked at several economic reasons that drive this phenomenon – MOs revealing private information and MOs' mechanical consumption of available liquidity – both leading to the change in the best-quoted price (immediate or subsequent). It is, thus, intuitive to explore a relationship of how the size of MO can influence the degree of impact. Therefore, we define $\mathcal{R}\left(\upsilon,1\right)$ as a volume dependent lag-1 impact response function and further explore its’ functional form.

\clearpage
\vspace{5mm}
\noindent
\textbf{Feature Engineering: Extraction and Selection}

\vspace{5mm}
\noindent
In this respect, our feature selection process throughout the study has been guided by economic intuition. That is, following a comprehensive study of subject literature we have identified features that we believe can help explain market impact; as opposed to obtaining the inferred relationships between the price impact and various observed market variables using feature extraction techniques.

\vspace{5mm}
\noindent
To arrive at a point where we can examine the impact response function, we have also engineered several features such as \textit{mid-price, spread, the standard deviation of price fluctuations, normalised trade volume} and the \textit{response function} itself. Whilst the former features are not going to be used directly for defining price impacts functional form, quantities like spread and standard deviation illustrate important qualities of our dataset which we highlighted in the previous section.

\vspace{5mm}
\vspace{5mm}
\noindent
\textbf{Normalised Volume}

\vspace{5mm}
\noindent
To account for daily trends and to be able to compare response function behaviour among stock with different average traded volume characteristics, we normalise the MO volume $\upsilon$ by the average volume at the opposite side best quote ${\bar{V}}_{best}$. The mean volume used for normalisation is recalculated each day ensuring the robustness of the relative scale.

\clearpage
\vspace{5mm}
\noindent
\textbf{Descriptive Statistics}

\vspace{5mm}
\noindent
For this exercise, we recorded the normalised volume of each MO and its corresponding lag-1 response function $\mathcal{R}\left(1\right)$. As a part of data pre-processing, we have removed outliers and aggregated average response per observed values of normalised volume. Jupyter Notebook (document that illustrates the code and its execution) \textit{Conditioning on Trade Volume} in Appendix E describes the detailed procedure. 

\vspace{5mm}
\noindent
An initial analysis of distribution of normalised volume, showed that the majority of trades for both large tick EBAY and small tick PCLN have a relatively small normalised volume (example in Figure 4). This means we have many more observations for smaller trade volumes than for trades of a larger size.

\vspace{5mm}
\vspace{5mm}
\begin{figure}[hbt!]
    \centering
    \includegraphics[height = 7cm, width=9cm]{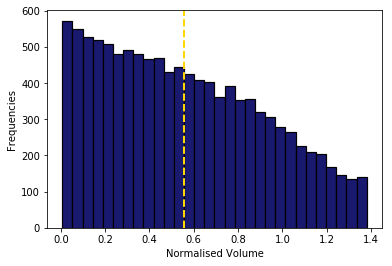}
    \caption{ Normalised volume distribution for PCLN. Yellow line representing the mean normilised volume.}
    \label{fig:normVOl}
\end{figure}

\noindent
To regularise the data, we ordered the series by normalised volume and then split the dataset in two: trades with normalised volume less than 0.1 and the rest of observations. Then we subsampled each partition at different frequencies.

\vspace{5mm}
\vspace{5mm}
\noindent
\textbf{Subsampling}

\vspace{5mm}
\noindent
The LOBSTER data represents the evolution of the LOB throughout each trading day sampled by the arrival of new events, as opposed to sampling at regular time intervals. In our study, we are similarly interested in the series of events as they arrive in the exchange, and more specifically MO executions. 

\vspace{5mm}
\noindent
In our initial analysis of market data, we conducted several data subsampling experiments where we regularised time series by sampling at regular intervals in several feature spaces. The main purpose of subsampling is to transform a series of observations that arrive at irregular intervals (such as LOBSTER data) into a homogeneous series. Such form is akin to a tabular representation and is often referred to as \textit{bars}.  In the Jupyter Notebook \textit{Create Bars} (see Appendix E) several standard bars are created including time, tick, volume and dollar bars. Volume bars, for example, sample data every time a certain threshold (number of shares) of the security has been exchanged. Similarly, other types of bars sample daily market data at regular intervals in tick, time and dollar dimensions. 

\vspace{5mm}
\noindent
To estimate the relationship between response function and trade volume, however, we examine every market order – its corresponding normalised volume and lag-1 response. Therefore, a slightly different kind of regularisation applies in this case: for trades with smaller trade volume we sample the data every 0.01 increment in the (ordered) series of normalised volume; for larger volume trades we sample the data every 0.1 increment in normalised volume observation. This approach to regularisation allows us to normalise the dataset according to the density of observations. An exact procedure is demonstrated in the Jupyter Notebook \textit{Conditioning on Trade Volume} which shows how we have visualised lag-1 response function conditioned on normalised trade volume (we collected the data during execution of \textit{uncond\_market\_impact.py} – available in Appendix E).

\clearpage
\vspace{5mm}
\noindent
\textbf{Lag-1 Response Function Conditioned on Normalised Trade Volume}

\vspace{5mm}
\noindent
From our earlier discussions it was intuitive to assume that larger MOs would affect the price more than smaller MOs. However, the empirical results in Figure 5  demonstrate that the effect of trade volume on the price is very faint: small tick stock PCLN exhibits almost constant monotonic relationship between normalised trade volume and the price impact (as measured by $\mathcal{R}\left(\upsilon\ ,1\right)$); whilst  large tick stock EBAY illustrates some functional dependency for small volumes but it dilutes and becomes monotonous as the normalised volume increases.

\vspace{5mm}
\begin{figure}[hbt!]
  \begin{subfigure}{7cm}
    \centering\includegraphics[width=7cm, height=5cm]{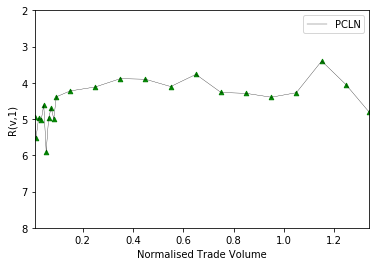}
  \end{subfigure}
  \begin{subfigure}{7cm}
    \centering\includegraphics[width=7cm, height=5cm]{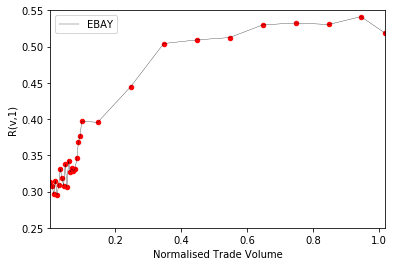}
  \end{subfigure}
  \caption{Lag-1 Response Function Conditioned on Normalised Trade Volume for PCLN (left) and EBAY (right)}
  \label{fig:lag1_response}
\end{figure}

\vspace{5mm}
\noindent
From this visual representation of lag-1 impact conditioned on the trade volume we note that the response function appears to be concave and even almost constant for small tick stocks such as PCLN. It would be necessary, however, to compare observations for a larger number of securities to confirm the robustness of such findings. We refer to the works of Bouchaud et al. (2018) and Zarinell et al. (2015) who obtained similar results. The former attribute concavity to a conditioning bias called selective liquidity taking. Appendix B.1 elaborates further on this phenomenon.

\vspace{5mm}
\subsubsection{Order-Flow Imbalance}
\vspace{5mm}
\noindent
An alternative approach to measuring impact of trades on price is to observe market impact not at trade-by-trade level, as we were doing before, but instead at \textit{aggregated order-flow} scale. In other words, in contrast to looking at the individual impact of trades, we aggregate multiple MOs’ signed volumes as a single \textit{order-flow imbalance} and measure how the price has changed before and after.  Recall $\varepsilon$ as order sign (positive meaning buy MO, and negative – sell MO), and $\upsilon$ as the volume of an individual MO, we denote order-flow imbalance $\Delta V$ as:

\begin{eqnarray}
\Delta V = \sum_{n\in[t,\ t+T)}{{\ \varepsilon}_n\upsilon_n}
\end{eqnarray}

\vspace{5mm}
\noindent
where $T > 0$ tells the effect of how many events (MOs) is measured by $\Delta V$.

\vspace{5mm}
\noindent
In this regard, $\Delta V$ indicates the proportions of buy and sell orders that arrive during a given period (as measured in the number of MOs observed). Positive $\Delta V$ means there were more buy orders (in volume) during the specified period (buy orders have positive sign and volume causing $\Delta V > 0$ ), therefore the price is expected to rise (for reasons we have discussed earlier). Contrastingly, negative $\Delta V$ implies more recent sell orders, hence the price is expected to decline. 
\vspace{5mm}
\noindent
We further study the \textit{aggregate impact} conditioned on this volume imbalance, defined as:

\begin{eqnarray}
\mathbb{R}(\Delta V,T) := \mathbb{E}[m_{t+T} - m_t|\Delta V] 
\end{eqnarray}

\vspace{5mm}
\noindent
$\mathbb{R}(\Delta V,T)$ measures change in the mid-price immediately before the original MO and after a sequence of $T$ MOs $(m_{t+T}$ being a mid-price immediately before the next MO after $T$ MOs).

\vspace{5mm}
\noindent
\textbf{Aggregate Response Function Conditioned on Trade-Flow Imbalance}

\vspace{5mm}
\noindent
We choose to investigate the relationship between aggregate impact (measured in  dollar cents) and the order-flow imbalance (measured in shares) over $T = 5,\ 10,\ 20,\ 50$ for TSLA stock. Our approach is demonstrated in the Jupyter Notebook \textit{Conditioning on Order-Flow Imbalance} in Appendix E.

\begin{figure}[hbt!]
  \begin{subfigure}{7cm}
    \centering\includegraphics[width=6.5cm, height=5cm]{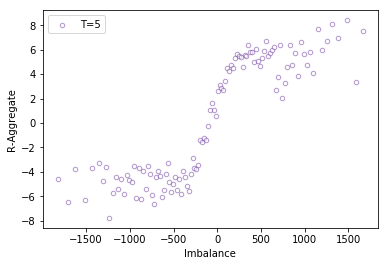}
  \end{subfigure}
  \begin{subfigure}{7cm}
    \centering\includegraphics[width=6.5cm, height=5cm]{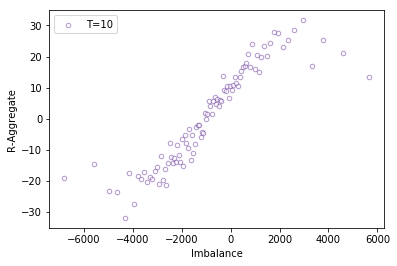}
  \end{subfigure}
  
  \begin{subfigure}{7cm}
    \centering\includegraphics[width=6.5cm, height=5cm]{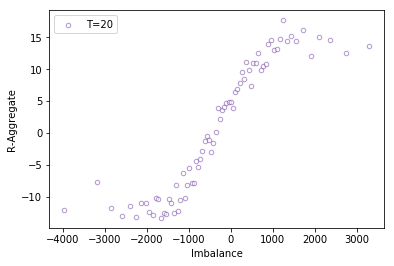}
  \end{subfigure}
  \begin{subfigure}{7cm}
    \centering\includegraphics[width=6.5cm, height=5cm]{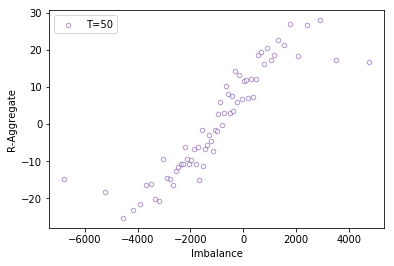}
  \end{subfigure}
  
  \caption{Aggregate impact (in dollar cents) of order-flow imbalance (in shares) (in over T=5 (top left), T=10 (top right), T=20 (bottom left) and T=50 (bottom right) for TSLA stock}
  \label{fig:order-imb}
\end{figure}

\vspace{5mm}
\noindent
In this part of the study we did not normalise (by corresponding average daily statistics, as done in the previous section when conditioning on trade volume) aggregate impact or imbalance, therefore, each experiments’ results are presented in individual subplot. The main purpose of this exercise was to visually understand the functional form of relationship between order-flow imbalance and the aggregated price response function. The following interesting properties can be noticed from Figure 6:

\begin{itemize}

    \item 	Imbalance amplitude increases with $T$ – as we aggregate the effect of more trades, the order-flow imbalance increases in absolute value as well. For instance, when observing the impact of $T = 5$ MOs, imbalance ranges from around -1500 to 1500 shares. However, once $T$ is increased to 10, imbalance scope expands to reaching 6000 shares in absolute value. Interestingly, the functional form of relationship seems the same for all $T$ despite different scales.
    
    \item 	For smaller $\Delta V$ the relationship with  $\mathbb{R}(\Delta V,T$) appears to be linear for all \break observed  $T = 5,\ 10,\ 20,\ 50$.
    
    \item 	With increase in $|\Delta V|$ (absolute value of order-flow imbalance) the relationship takes on a more concave (rather than linear) form, similar to our previous analysis of lag-1 response function conditioned on normalised volume. Again, this may be attributed to the selective liquidity bias, as described in Appendix B.1.
    
\end{itemize}

\subsection{Results}
\vspace{5mm}
\noindent
In the previous section we stated that the dependency between order-flow imbalance and corresponding aggregate price impact appears linear for smaller absolute values of $\Delta V$. In this segment, we conduct statistical experiments, as outlined in the research methodology section, to quantitatively infer functional form of market impact conditional on order-flow imbalance.  Our approach is illustrated in the Jupyter Notebook \textit{Market Impact Functional Form} in Appendix E.

\vspace{5mm}
\subsubsection{Functional Form of Market Impact}
\noindent

\vspace{5mm}
\noindent
\textbf{Descriptive Statistics}

\vspace{5mm}
\noindent
We use the data collected for TSLA stock over the entire year of 2015 with order-flow imbalance observed over $T = 10$ MOs with the total of 47,473 observations of $\Delta V$ and $\mathbb{R}(\Delta V,T$). We remove the outliers beyond 3 standard deviations and subsample the data to reduce the \textit{noise}. Visually our dataset is described by Figure 7. We also illustrate the distributions of $\mathbb{R}(\Delta V,T)$ – aggregate impact, and $\Delta V$ – order-flow imbalance in Figure 8.

\vspace{5mm}
\vspace{5mm}
\begin{figure}[hbt!]
    \centering
    \includegraphics[height = 7cm, width=9cm]{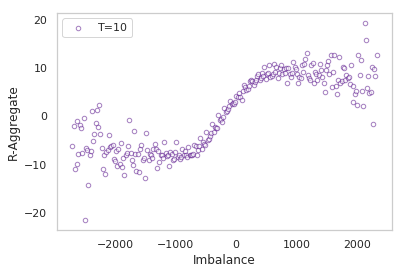}
    \caption{Aggregate impact (in dollar cents) conditioned on order-flow imbalance (in shares) for TSLA stock during 2015 after data pre-processing}
    \label{fig:TSLA-T10}
\end{figure}

\vspace{8mm}
\noindent
\begin{figure}[hbt!]
  \begin{subfigure}{7cm}
    \centering\includegraphics[width=6.5cm, height=5cm]{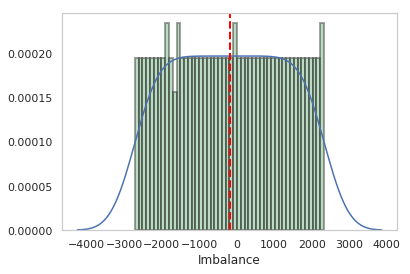}
  \end{subfigure}
  \begin{subfigure}{7cm}
    \centering\includegraphics[width=6.5cm, height=5cm]{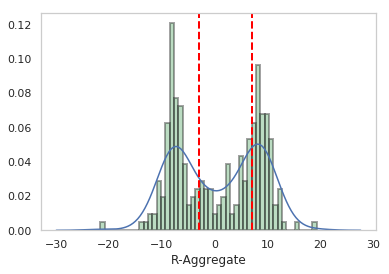}
  \end{subfigure}
  
  \caption{Aggregate impact (in dollar cents) and order-flow imbalance (in shares) distributions for TSLA during 2015. Red lines representing means - for aggregate response two lines are positive and negative means.}
  \label{fig:TSLA-imb-ag}
\end{figure}

\vspace{5mm}
\noindent
These are in accordance with the previous illustration – imbalance distribution is uniform, meaning that asymmetry (order-flow imbalance) of every size on either direction of trade is equally likely. Aggregate impact, on the other hand, seems to have more observations for a specific absolute value of around 10 dollar cents. This concentration of aggregated price impact, again, can be explained by the selective liquidity bias - larger orders (therefore order-flow imbalance) only consume the liquidity available at the best opposite-side quote, hence, the impact does not increase beyond certain margin (is concave).

\vspace{5mm}
\noindent
We further explore the correlation between the two observed variables and calculate the correlation coefficient $\rho = 0.8538$ which indicates a strong positive relationship between order-flow and price impact (i.e., the greater the imbalance the bigger the market impact).  

\vspace{5mm}
\noindent
The summary of other statistics for order-flow imbalance and aggregate impact is presented in Appendix C - Table 9.

\vspace{5mm}
\noindent
\textbf{Supervised Linear Regression}

\vspace{5mm}
\noindent
We first estimate functional relationship between $\Delta V$ and $\mathbb{R}(\Delta V,T)$ using a \textit{Linear Regression} model which employs the OLS approach.

\vspace{5mm}
\noindent
We split our dataset randomly into two partitions: the \textit{training data} to fit the model, and the \textit{test data} reserved to verify predictive accuracy of our model on unseen data. To quantify this accuracy, Mean Squared Error (MSE) is calculated to assess how values estimated by the regression model differ from the true observations in the test dataset.

\vspace{5mm}
\noindent
In the case of Linear Regression, the MSE is 3.86 meaning that, on average, our predicted aggregated price impact was either above or below the true value by 3.86 dollar cents. We graphically illustrate predictive linear model and true observations in the Figure 9.

\clearpage

\vspace{5mm}
\noindent
\textbf{Decision Tree}

\vspace{5mm}
\noindent
As an alternative to Linear Regression, we estimate a model using \textit{Decision Tree Regression} – a supervised learning technique that predicts values of regressand by learning decision rules derived from observations. In contrast to Linear Regression, Decision Tree model is not linear in parameters, therefore, it utilises MLE for parameter estimation. Similarly, we train our model on one partition of the data and test it on another unseen dataset. The MSE for Decision Tree model is 2.5 dollar cents – a better estimate than Linear Regression model. A visual representation of predicted and observed values can be seen in Figure 9.

\vspace{5mm}
\noindent
\begin{figure}[hbt!]
  \begin{subfigure}{7cm}
    \centering\includegraphics[height = 6cm, width=7cm]{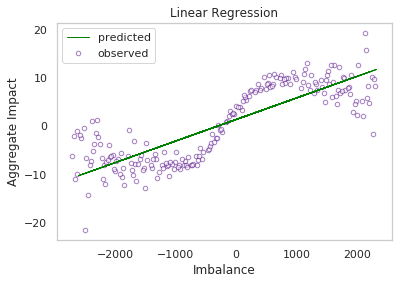}
  \end{subfigure}
  \begin{subfigure}{7cm}
    \centering\includegraphics[height = 6cm, width=7cm]{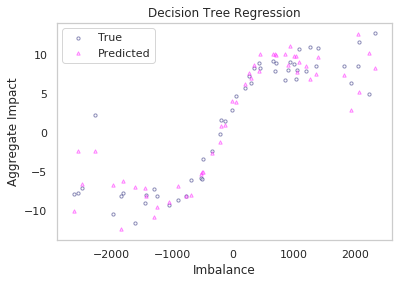}
  \end{subfigure}
  
  \caption{Linear Regression model and true observations of market impact for TSLA (left). Decision Tree model and true observations of market impact for TSLA (right) (Decision Tree Regression only displaying testing observations).}
  \label{fig:LR-DT1}
\end{figure}

\clearpage
\noindent
\textbf{Cross-Validation: Model Selection}

\vspace{5mm}
\noindent
Splitting the dataset into two partitions – one for training and one for testing – is a naïve approach to testing predicting accuracy. Thus, we employ cross-validation to determine which model produces better estimates.

\vspace{5mm}
\noindent
Cross-validation is conducted using a helper function that splits the data into training and test sets, fits the model and calculates scores (MSE and $R2$ in our case) for a number of consecutive runs. We perform cross-validation using 10 partitions on both models. Figure 10 demonstrate MSEs and $R2$ coefficients for Linear Regression and Decision Tree Regression (Table 11 in Appendix C shows the exact details).

\vspace{5mm}
\noindent
\begin{figure}[hbt!]
  \begin{subfigure}{7cm}
    \centering\includegraphics[height = 5.5cm, width=7cm]{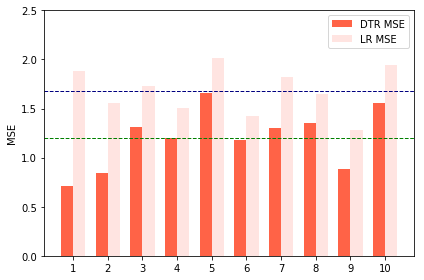}
  \end{subfigure}
  \begin{subfigure}{7cm}
    \centering\includegraphics[height = 5.5cm, width=7cm]{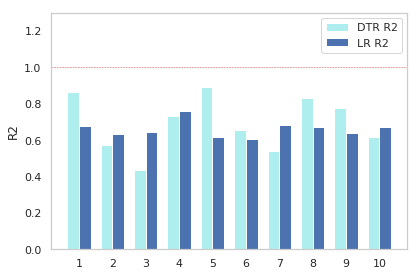}
  \end{subfigure}
  
  \caption{MSE (left) with average MSE for Linear Regression in blue, and average MSE for Decision Tree Regression in green; and R2 results (right) for Linear Regression and Decision Tree Regression.}
  \label{fig:MSER2-1}
\end{figure}

\vspace{5mm}
\noindent
From the two figures it is evident that the Decision Tree Regression displayed better predictive performance - its average MSE is slightly lower than MSE of Linear Regression model; and the coefficient $R2$ (proportion of variance in the observed data that is explained by the estimated model) is higher for Decision Tree regression as well. However, the results of cross-validation illustrate that Decision Tree Regression does not always produce more accurate predictions. Despite this, we strongly believe the performance of ML algorithm could be improved in future studies by exploring various other features, learning models and tuning the parameters.

\vspace{5mm}
\subsubsection{Kyle's Lambda}
\noindent


\vspace{5mm}
\noindent
The functional linearity between price impact and order-flow imbalance is an empirical illustration of Kyle's model which was introduced in the earlier Literature Review chapter. As we recall, Kyle stated that the price adjustment must be linear in total signed volume, expressed as: 
\begin{eqnarray}
\Delta p = \lambda \varepsilon V
\end{eqnarray}
\noindent
Therefore, the slope of linear relationship that we have observed for smaller volume imbalances is, what's known as, a \textit{Kyle's lambda}. 

\vspace{5mm}
\noindent
\textbf{Model Estimation}

\vspace{5mm}
\noindent
In order to estimate the slope of this linear region of $\mathbb{R}(\Delta V,T)$, we train the model specifically on the subsection of our observations where the relationship appears linear (as illustrated in the Figure 11).

\vspace{5mm}
\noindent
\begin{figure}[hbt!]
  \centering\includegraphics[height = 6cm, width=8cm]{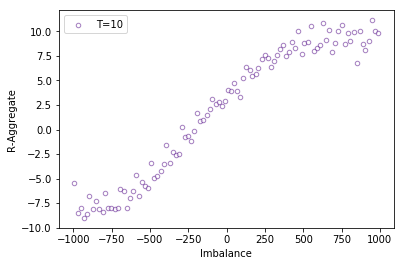}
  \caption{Subsection of observation that demonstrate linearity between order-flow imbalance and price impact for TSLA in 2015.}
\end{figure}

\vspace{5mm}
\noindent
Again, we use Linear Regression and Decision Tree Regression to fit the model and cross-validate it. The models and corresponding statistics about prediction accuracy (MSE and $R2$) are illustrated in the Figure 12.

\begin{figure}[hbt!]
  \begin{subfigure}{7cm}
    \centering\includegraphics[width=6.5cm, height=5cm]{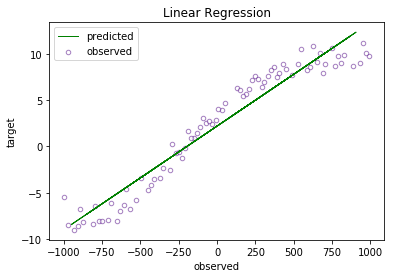}
  \end{subfigure}
  \begin{subfigure}{7cm}
    \centering\includegraphics[width=6.5cm, height=5cm]{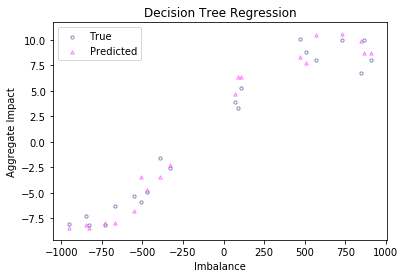}
  \end{subfigure}
  
  \begin{subfigure}{7cm}
    \centering\includegraphics[width=6.5cm, height=5cm]{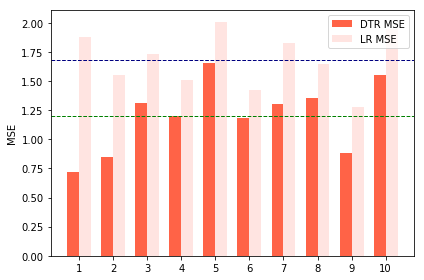}
  \end{subfigure}
  \begin{subfigure}{7cm}
    \centering\includegraphics[width=6.5cm, height=5cm]{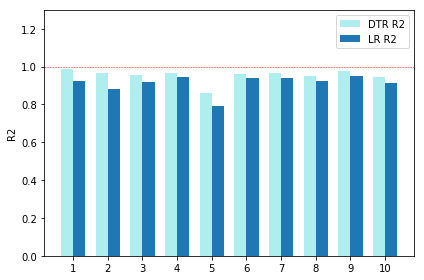}
  \end{subfigure}
  
  \caption{Top: models of Linear and Decision Tree Regressions for predicting market impact conditioned on order-flow imbalance (Decision Tree Regression only displaying testing observations); bottom: measures of goodness of fit for the two models (left: green line - average MSE for Decision Tree, blue line - average MSE for Linear Regression}
  \label{fig:linear}
\end{figure}

\vspace{5mm}
\noindent
In this instance, both models have a good performance according to $R2$ coefficient (both Linear and Decision Tree Regressions seem to explain much of the variance in the observed data); however, Decision Tree Regression model demonstrates much lower average MSE by almost 0.50 dollar cents. This supports our earlier remark about the potential for more accurate ML predictive models. 

\newpage
\noindent
\textbf{Measure of Market Liquidity}

\vspace{5mm}
\noindent
The slope of our linear model, the Kyle's $\EuScript{\lambda}$, was estimated as 0.011. This is regarded as the measure of (il-)liquidity of a market. For TSLA, we can postulate that the market is quite liquid, since the value of $\EuScript{\lambda}$ is low. However, for more robust conclusions it is necessary to compare measurements of $\EuScript{\lambda}$ across multiple stocks and venues.
\section{Conclusion and Future Work}
\vspace{5mm}

\subsection{Summary}

\vspace{5mm}
\noindent
The focus of this research was to find a functional form of market impact using parametric machine learning algorithms. We began with an overview of the financial market microstructure  and the role of information (and more importantly information asymmetry) in the price formation (discovery) process. We then looked at a number of models of the information efficiency in the market, and how agents' interactions influence trading. Naturally we arrived at the definition of market impact - the focal point of this study. 

\vspace{5mm}
\noindent
Following suit, we used trades and quotes dataset from NASDAQ electronic exchange to measure unconditional market impact, price impact conditioned on normilised trade volume and the aggregate price response as a function of order-flow imbalance. We have learnt that the unconditional price impact of market orders is proportional to the average spread for the stock. Moreover, we observed no functional dependency between price response function and normilised trade volume - partially attributed to the selective liquidity bias. In our analysis of aggregate price impact conditioned on the order-flow imbalance, we discerned sublinear relationship between these two observed variables. 

\vspace{5mm}
\noindent
Consequently, we employed Linear Regression and Decision Tree Regression models to quantitatively measure the functional form of such relationship. We used the two models to predict values of aggregated price impact of order-flow imbalance after a number of consecutive market orders. Decision Tree Regression - a machine learning algorithm that is not linear in parameters, illustrated better predictive accuracy than traditional Linear Regression, during cross-validation.

\vspace{5mm}
\subsection{Future Work}

\vspace{5mm}
\noindent
We identify a number of ways in which our study can be further developed. The below list provides  a summary of our suggestions. These are further explained in the Appendix D.

\begin{itemize}

    \item To explore the statistical and practical application of ML models in estimation of market impact it would be necessary to conduct a comparison with performance of current standard (for academics and practitioners) measures of price impact (i.e., Volume Weighted Average Price, VWAP); as well as other theoretical and empirical formulas of market impact.
    
    \item Alternative data sources can be used for study of the impact of meta-orders (see Appendix D.1.1).
    
    \item It would be of acute importance to study liquidity dynamics and impact of trades in illiquid markets - for example, in stressed market conditions (see Appendix D.1.2).
    
    \item Furthermore, the study can be expanded to other financial products (not just equities) (see Appendix D.1.3).
    
    \item Accurate measure of liquidity (thus market impact) facilitates the understanding of optimal execution, further contributing to the study of optimal trading strategies (see Appendix D.1.4).
    
    \item Experimental part of this research has been computationally intensive - it would be beneficial to adopt high-performance computational resources when analysing high-frequency data (see Appendix D.1.5).
    
\end{itemize}

\clearpage
\addcontentsline{toc}{section}{References}
\nocite{*}
\printbibliography

@misc{Hua11,
	title = {LOBSTER: Limit Order Book Reconstruction System},
	url = {http://dx.doi.org/10.2139/ssrn.1977207},
	month = {06},
	year = {2011},
	label = {Hua11},
	author = { Ruihong Huang and Tomas Polak},
}

@book{Har03,
	location = {New York},
	title = {Trading and Exchanges: Market Microstructure for Practitioners},
	year = {2003},
	label = {Har03},
	publisher = {Oxford University Press},
	author = { Larry Harris },
}

@article{Ach06,
	pages = {460-463},
	title = {Liquidity Risk: Causes, Consequences and Implications for Risk Management},
	volume = {41},
	issue = {6},
	year = {2006},
	label = {Ach06},
	journal = {Economic and Political Weekly},
	author = { Viral Acharya },
}

@misc{Blo16,
	title = {Bloomberg Liquidity Assessment LQA},
	url = {https://data.bloomberglp.com/professional/sites/10/LQA-Fact-Sheet-1.pdf},
	month = {06},
	year = {2016},
	label = {Blo16},
	author = { Bloomberg},
}

@article{Bru09,
	pages = {2201-2238},
	title = {Market Liquidity and Funding Liquidity},
	volume = {22},
	issue = {6},
	year = {2009},
	label = {Bru09},
	journal = {Review of Financial Studies},
	author = { Markus K.Brunnermeier and Lasse Heje Pedersen },
}

@book{Dav08,
	location = {New York},
	title = {Understanding Risk: The Theory and Practice of Financial Risk Management},
	year = {2008},
	label = {Dav08},
	publisher = {Chapman and Hall/CRC},
	author = { David Murphy },
}

@misc{Placeholder1,
	title = {Risk.net},
	url = {https://www.risk.net/asset-management/4644191/quants-turn-to-machine-learning-to-model-market-impact},
	month = {05},
	year = {2017},
	label = {Placeholder1},
	author = { Sebastian Day },
}

@article{Placeholder2,
	pages = {1-16},
	title = {The Market Impact Puzzle},
	year = {2018},
	label = {Placeholder2},
	journal = {SSRN Electronic Journal},
	author = { Albert S. Kyle and Anna A. Obizhaeva },
}

@article{Placeholder3,
	year = {2016},
	title = {Predicting Market Impact Costs Using Nonparametric Machine Learning Models},
	label = {Placeholder3},
	journal = {PLoS ONE 11(2):e0150243 - DOI: 10.1371/journal.pone.0150243},
	author = { Saerom Park and Jaewook Lee and Youngdoo Son },
}

@article{Con10,
	pages = {1-26},
	title = {The Price Impact of Order Book Events},
	year = {2010},
	label = {Con10},
	journal = {Journal of Financial Econometrics},
	author = { Roma Cont and Arseniy Kukanov and Sasha Stoikov },
}

@article{Placeholder4,
	pages = {257-275.},
	title = {Market Microstructure.},
	year = {1976},
	label = {Placeholder4},
	journal = {Journal of Financial Economics},
	author = { Mark Garman },
}

@article{Kyl85,
	pages = {1315–1335},
	title = {Continuous auctions and insider trading},
	volume = {vol. 53},
	issue = {6},
	year = {1985},
	label = {Kyl85},
	journal = {Econometrica},
	author = { Albert S. Kyle},
}

@article{Gro80,
	pages = {573-585},
	title = {On the Efficiency of Competitive Stock Markets Where Trades Have Diverse Information},
	volume = {31},
	issue = {2},
	year = {1976},
	label = {Gro80},
	journal = {The Journal of Finance},
	author = { Sanford J. Grossman},
}

@article{Ber98,
	pages = { 1-50},
	title = {Optimal control of execution costs},
	volume = {1},
	issue = {1},
	year = {1998},
	label = {Ber98},
	journal = {Journal of Financial Market},
	author = { Dimitris Bertsimas and Andrew W. Lo},
}

@article{Alm99,
	pages = {61-63},
	title = {Value Under Liquidation},
	volume = {12},
	year = {1999},
	label = {Alm99},
	journal = {Risk},
	author = { Robert Almgren and Neil Chriss },
}

@article{Alm01,
	pages = {5-39},
	title = {Optimal Execution of Portfolio Transactions},
	volume = {3},
	year = {2001},
	label = {Alm01},
	journal = {Journal of Risk},
	author = { Robert Almgren and Neil Chriss },
}

@article{Ave08,
	pages = {217-224},
	title = {High Frequency Trading in a Limit Order Book},
	volume = {8},
	issue = {3},
	year = {2008},
	label = {Ave08},
	journal = {Quantitative Finance},
	author = { Marco Avellaneda and Sasha F. Stoikov }
}

@article{Gue13,
	pages = {477–507},
	title = {Dealing with the Inventory Risk. A solution to the market making problem},
	volume = {7},
	issue = {4},
	year = {2013},
	label = {Gue13},
	journal = {Mathematics and Financial Economics},
	author = {  Olivier Guéant and Charles-Albert Lehalle  and Joaquin Fernandez Tapia },
}

@book{Placeholder5,
	year = {2018},
	title = {Trades, Quotes and Prices Financial Markets Under the Microscope},
	label = {Placeholder5},
	publisher = {Cambridge University Press},
	author = { Jean-Philippe Bouchaud and Julius Bonart  and Jonathan Donier and Martin Gould },
}

@article{Gou13,
	pages = {1709-1742 },
	title = {Limit order books},
	volume = {13},
	issue = {11},
	year = {2013},
	label = {Gou13},
	journal = {Quantitative Finance},
	author = { Martin D. Gould and Mason A. Porter and  Stacy Williams and Mark Mcdonald and Daniel J. Fenn and Sam D. Howison},
}

@article{Bon17,
	pages = {1601-1616},
	title = {Latency and liquidity provision in a limit order book},
	volume = {17},
	issue = {10},
	year = {2017},
	label = {Bon17},
	journal = {Quantitative Finance},
	author = { Julius Bonart and Martin Gould},
}

@book{Oli16,
	location = {New York},
	title = {The Financial Mathematics of Market Liquidity From Optimal Execution to Market Making},
	year = {2016},
	label = {Oli16},
	publisher = {Chapman and Hall/CRC},
	author = { Olivier Guéant },
}

@article{Smi03,
	pages = {481-514},
	title = {Statistical theory of the continuous double auction},
	volume = {3},
	issue = {6},
	year = {2003},
	label = {Smi03},
	journal = {Quantitative Finance},
	author = { Eric Smith and Doyne J. Farmer and László Gillemot and Supriya Krishnamurthy },
}

@article{Con101,
	pages = {549-563},
	title = {A Stochastic Model for Order Book Dynamics},
	volume = {58},
	issue = {3},
	year = {2010},
	label = {Con101},
	journal = {Operations Research},
	author = { Roma Cont and Sasha Stoikov and Rishi Talreja },
}

@article{Far05,
	pages = {2254-2259},
	title = {The predictive power of zero intelligence in financial markets},
	volume = {102},
	issue = {6},
	year = {2005},
	label = {Far05},
	journal = {Proceedings of the National Academy of Sciences of the United States of America},
	author = { Doyne J. Farmer and Paolo Patelli and Ilija J. Zovko },
}

@article{Hua15,
	pages = {107-112},
	title = {Simulating and Analyzing Order Book Data: The Queue-Reactive Model},
	volume = {110},
	issue = {509},
	year = {2015},
	label = {Hua15},
	journal = {Journal of the American Statistical Association},
	author = { Weibing Huang and Charles-Albert Lehalle and Mathieu Rosenbaum },
}

@book{Car15,
	location = {Cambridge},
	title = {Algorithmic and High-Frequency Trading},
	year = {2015},
	label = {Car15},
	publisher = {Cambridge University Press},
	author = { Alvaro Cartea and Sebastian Jaimungal and Jose Penalva },
}

@article{Eis12,
	pages = {1395-1419},
	title = {The price impact of order book events: market orders, limit orders and cancellations},
	volume = {12},
	issue = {9},
	year = {2012},
	label = {Eis12},
	journal = {Quantitative Finance},
	author = { Zoltan Eisler and Jean-Philippe Bouchaud and Julien Kockelkoren },
}

@book{Has07,
	year = {2007},
	title = {Empirical Market Microstructure: The Institutions, Economics, and Econometrics of Securities Trading},
	label = {Has07},
	publisher = {Oxford University Press },
	author = { Joel Hasbrouck },
}

@book{Mic11,
	year = {2011},
	title = {Investments: Principles of Portfolio and Equity Analysis},
	label = {Mic11},
	publisher = {Wiley},
	author = { McMillan Michael and Jerald E. Pinto and  Wendy L. Pirie and Gerhard Van de Venter and Lawrence E. Kochard },
}

@article{Fin76,
	pages = {1141-1148},
	title = {INSIDERS AND MARKET EFFICIENCY},
	volume = {31},
	issue = {4},
	year = {1976},
	label = {Fin76},
	journal = {The Journal of Finance},
	author = { Joseph E. Finnerty  },
}

@article{Sey86,
	title = {Insiders' profits, costs of trading, and market efficiency},
	volume = {16},
	issue = {2},
	year = {1986},
	label = {Sey86},
	journal = {Journal of Financial Economics },
	author = { Nejat H. Seyhun },
}

@article{Glo85,
	pages = {71-100},
	title = {Bid, ask and transaction prices in a specialist market with heterogeneously informed traders},
	volume = {14},
	issue = {1},
	year = {1985},
	label = {Glo85},
	journal = {Journal of Financial Economics },
	author = { Lawrence R. Glosten and Paul R. Milgrom },
}

@article{Mas14,
	pages = {042805},
	title = {Agent-based models for latent liquidity and concave price impact},
	volume = {89},
	issue = {4},
	year = {2014},
	label = {Mas14},
	journal = {Physical review E},
	author = { Iacopo Mastromatteo and  Bence Toth and Jean-Philippe Bouchaud},
}

@article{Don15,
	pages = {1109-1121},
	title = {A fully consistent, minimal model for non-linear market impact},
	volume = {15},
	issue = {7},
	year = {2015},
	label = {Don15},
	journal = {Quantitative Finance},
	author = { Donier Jonathan Bonart Julius Mastromatteo Iacopo Bouchaud Jean-Philippe },
}

@article{Bou09,
	pages = {57–160},
	title = {How markets slowly digest changes in supply and demand},
	year = {2009},
	label = {Bou09},
	journal = {Handbook of Financial Markets: Dynamics and Evolution},
	author = {Jean-Philippe Bouchaud and Doyne J. Farmer and Fabrizio Lillo },
}

@book{Leh18,
	year = {2018},
	title = {Market Microstructure in Practice},
	label = {Leh18},
	publisher = {World Scientific Publishing},
	author = {Charles-Albert Lehalle and Sophie  Laruelle },
}

@article{Bou091,
	year = {2009},
	title = {Price Impact},
	label = {Bou091},
	journal = {Encyclopedia of Quantitative Finance; arXiv:0903.2428 [q-fin.TR]; arXiv:0903.2428v1 [q-fin.TR},
	author = {Jean-Philippe Bouchaud},
}

@article{Mor09,
	year = {2009},
	title = {Market impact and trading profile of large trading orders in stock markets},
	label = {Mor09},
	journal = {arXiv:0908.0202 [q-fin.TR]; arXiv:0908.0202v1 [q-fin.TR]},
	author = {Magro E.  Moro and Javier  Vicente  and Luís G. Moyano and Gerig  Austin and J. Doyne Farmer and Gabriella Vaglica and Fabrizio Lillo and Rosario N. Mantegna},
}

@article{Alm05,
	pages = {57},
	title = {Direct Estimation of Equity Market Impact},
	year = {2005},
	label = {Alm05},
	journal = {Journal of Risk},
	author = {Robert Almgren and  Chee Thum and  Emmanuel Hauptmann and  Hong Li},
}

@article{Kyl89,
	pages = {317-355},
	title = {Informed Speculation with Imperfect Competition},
	volume = {56},
	issue = {3},
	year = {1989},
	label = {Kyl89},
	journal = {The Review of Economic Studies},
	author = {Albert S. Kyle },
}

@article{Bou04,
	pages = {176-190},
	title = {Fluctuations and response in financial markets: the subtle nature of `random' price changes},
	volume = {4},
	issue = {2},
	year = {2003},
	label = {Bou04},
	journal = {Quantitative Finance},
	author = { Jean-Philippe Bouchaud and Yuval Gefen  and Marc Potters  and  Matthieu Wyart},
}

@misc{Hau11,
	title = {Limit Order Flow, Market Impact and Optimal Order Sizes: Evidence from NASDAQ TotalView-ITCH Data.},
	year = {2011},
	label = {Hau11},
	publisher = {SSRN},
	author = {Nikolaus Hautsch  and  Ruihong Huang},
}

@article{Obi13,
	pages = {1-32},
	title = {Optimal trading strategy and supply/demand dynamics},
	volume = {16},
	issue = {1},
	year = {2013},
	label = {Obi13},
	journal = {Journal of Financial Markets},
	author = { Anna A. Obizhaeva and Jiang  Wang},
}

@article{Zar15,
	pages = {2424-8037},
	title = {Beyond the square root: Evidence for logarithmic dependence of market impact on size and participation rate},
	volume = {1},
	issue = {2},
	year = {2015},
	label = {Zar15},
	journal = {Market Microstructure and Liquidity},
	author = {Elia Zarinelli and Michele Treccani and J. Doyne Farmer and  Fabrizio Lillo},
}

@article{Web05,
	pages = {357-364 },
	title = {Order book approach to price impact},
	volume = {5},
	issue = {4},
	year = {2005},
	label = {Web05},
	journal = {Quantitative Finance},
	author = {P. Weber B. Rosenow},
}

@book{Tor97,
	location = {Berkeley},
	title = {Market Impact Model: Handbook},
	year = {1997},
	label = {Tor97},
	publisher = {Barra Inc.},
	author = { N. Torre },
}

@article{Loe83,
	pages = {39-44},
	title = {Trading Cost: The Critical Link between Investment Information and Results},
	volume = {39},
	issue = {3},
	year = {1983},
	label = {Loe83},
	journal = {Financial Analysts Journal},
	author = {Thomas F. Loeb},
}

@article{Kyl16,
	pages = {1345-1404},
	title = {Market Microstructure Invariance: Empirical Hypotheses},
	volume = {84},
	issue = {4},
	year = {2016},
	label = {Kyl16},
	journal = {Econometrica},
	author = { Albert  S. Kyle and Anna A. Obizhaeva},
}

@book{Mit97,
	location = {New York},
	title = {Machine Learning},
	year = {1997},
	label = {Mit97},
	publisher = {McGraw-Hill, Inc},
	author = {Thomas M. Mitchell},
}

@book{Jan181,
	year = {2018},
	title = {Hands-On Machine Learning for Algorithmic Trading},
	label = {Jan181},
	publisher = { Packt Publishing},
	author = {Stefan Jansen},
}

@article{Alm03,
	pages = {1-18},
	title = {Optimal execution with nmiscar impact functions and trading-enhanced risk},
	volume = {10},
	issue = {1},
	year = {2003},
	label = {Alm03},
	journal = {Applied Mathematical Finance},
	author = {Robert F. Almgren},
}

@book{Mur12,
    year = {2012},
	title = {Machine Learning: A Probabilistic Perspective},
	label = {Mur12},
	publisher = {MIT Press},
	author = {Kevin P. Murphy  },
	
}

@misc{reuters14,
	title = {What's behind the global stock market selloff?},
	url = {https://www.reuters.com/article/us-usa-markets-idUSKCN0VL0XO},
	month = {02},
	year = {2014},
	label = {reuters14},
	author = {David Randall and David Gaffen},
}

\clearpage
\appendix
\appendixpage
\begin{appendices}
\section{}

\subsection{\textit{Market and Funding Liquidity}}

\vspace{5mm}
\noindent
As outlined by Brunnermeier and Pedersen (2009), the market and funding liquidity concepts are mutually reinforcing. Traders provide market liquidity, and their ability to do so is dependent on the availability of funds. Conversely, traders funding needs, such as margin and capital requirements (i.e., collateral needed to enter into a trade) depend on the market liquidity of the asset. More accurately, when funding liquidity becomes scarce, traders become reluctant to take on capital intensive positions, lowering market liquidity and increasing margin and capital requirements. This increased risk, and thus cost of financing a trade (\textit{margin} or \textit{haircut}) is thus dependent on efficient streams of market liquidity.

\section{}
\subsection{\textit{Selective Liquidity Taking Under Market Fragmentation}}

\vspace{5mm}
\noindent
An imperative consequence of information asymmetry that is crucial to understanding of how markets operate is that even highly liquid markets are in fact not sufficiently liquid after all Bouchaud et al. (2018). That is, to minimise the risk of adverse selection, MMs providing liquidity operate in a regime of small revealed liquidity, and large latent liquidity. Consequently, the available liquidity at any given moment only reflects a small fraction of the potentially huge underlying supply and demand.  This is often captured by general equilibrium models such as Kyle (1985), Mastromatteo, Toth and Bouchaud (2014), which are based on interactions between rational agents who take optimal decisions.

\vspace{5mm}
\noindent
The scarcity of liquidity has an immediate and important consequence: large trades, or \textit{meta-orders}, must be fragmented into smaller \textit{child orders} that can be slowly digested by markets. Empirical observations such as Bouchaud (2009) suggest that order splitting is a very common practice in a wide range of markets.  In recent years, it has become common to trade assets on several different electronic platforms simultaneously. This increased fragmentation means it is necessary to split an order not  just through time, but also through space - to selectively consume liquidity (submit trades no larger than the available liquidity at the corresponding price) across all available venues (Lehalle and Laruelle, 2018).

\vspace{5mm}
\noindent
In the earlier discussions of functional form of relationship between response function and normalised volume, as well as, dependency of aggregate impact on order-flow imbalance, we have attributed concavity to selective liquidity taking.  This could be partially explained by the fact that the larger orders arrive only when there is a large volume available at the opposite side best quote. In other words, larger volume trades do not cause bigger impact because they only consume available liquidity and no more. Therefore, the price impact remains monotonic (concave) for different size trades. 

\vspace{5mm}
\noindent
The study of meta-orders and hidden orders (hidden in the sense that the true order size is not made public to minimize information leakage) yields surprising empirical findings for the functional form of market impact. However, the existing empirical literature on the subject is limited, with Torre (1997), Almgren (2003), Moro et al. (2009), being the few publications in this area.  This is attributed to the difficulty in obtaining access to data that could facilitate statistical reconstruction of transactions against trading account member codes. This would otherwise permit us to identify the parent meta-orders to which a child belongs to. The limitation of our study stems from this lack of proprietary datasets which clearly identifies the source of large orders. For this reason, latent liquidity and the impact of meta-orders are beyond the scope of this project.

\vspace{5mm}
\begin{figure}[hbt!]
    \centering
    \includegraphics[height = 7cm, width=14.5cm]{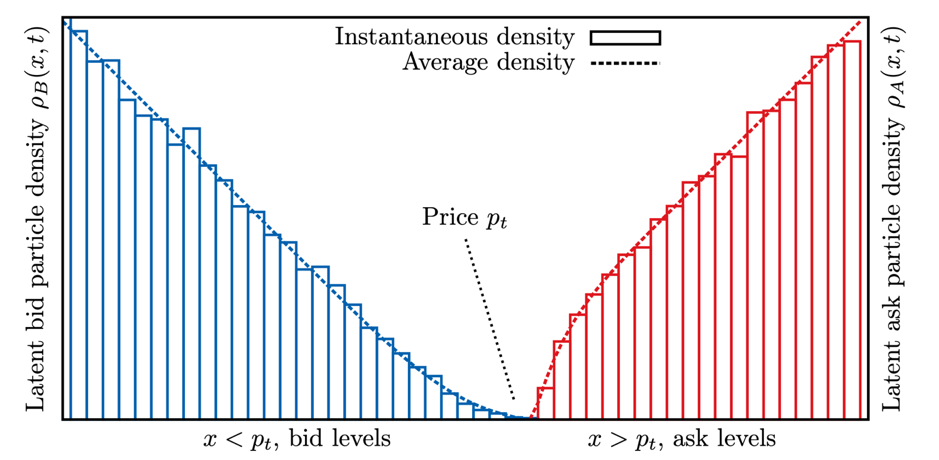}
    \caption{Latent Order Book in the presence of a meta-order, with bid orders (blue boxes) and ask orders (red boxes) sitting on opposite sides of the price line and subject to a stochastic evolution. (DONIER, Jonathan et al., 2015)}
    \label{fig:latentOrderBook}
\end{figure}

\subsection{\textit{Limit Orderbooks and Their Models}}

\vspace{5mm}
\noindent
The LOB keeps a record of order-flow events – the execution of trades (aggressive market orders) and the provision of liquidity (passive limit orders). These orders are managed by a matching engine in accordance with some well-defined algorithm (i.e., price time priority or pro-rata rules). The unusually rich, detailed, and high-quality historic LOB data facilitates the ability to analyse complex global phenomena that emerge as a result of the local interactions between many heterogeneous agents. 

\vspace{5mm}
\noindent
The earliest models of LOB generally depicted the evolution of order flow according to simple stochastic processes with set parameters. Smith et al. (2003) introduced a model in which limit order queues, as well as counter executed market orders and cancellations, transpire as mutually exclusive Poisson processes with fixed-rate parameters. The model has since been extended by Cont, Stoikov and Talreja (2010), who account for varying cancellation and limit order rates as a function of fluctuating prices. 

\vspace{5mm}
\noindent
Though these so-called \textit{zero intelligence} models (models in which order flows are assumed to be governed by specified stochastic processes) of LOBs perform relatively well at depicting some long-run statistical properties of genuine LOBs (see, e.g., Farmer, Patelli and Zovko (2005)); they are largely hindered by their exclusion of strategic considerations in which liquidity providers might adapt their flows in response to the actions of others. This constraint was recently alleviated in the work of Huang, Lehalle and Rosenbaum (2015). By incorporating an agent's strategic behaviour, they provide a more realistic picture of empirically observed facts about the price discovery and price formation processes, as well as, the general market quality.

\section{}

\subsection{\textit{The NASDAQ}}

\vspace{5mm}
\noindent
NASDAQ stands for “National Association of Securities Dealers Automated Quotation” and was originally founded in 1971 in New York, USA. The initial idea behind NASDAQ stock exchange was to give dealers the ability to post quotes electronically, thus making the process of selling smaller stocks (previously sold OTC) more efficient. This electrification of the trading process proved hugely successful. Today NASDAQ is the second-largest electronic exchange by market capitalisation in the world - currently exceeding 10 trillion US dollars. And the first by overall trading volume. 

\vspace{5mm}
\noindent
Unlike other major exchanges such as NYSE, NASDAQ does not have a physical location – it is an entirely digital marketplace.  NASDAQ normal trading hours are between 9:30 am and 4:00 pm Eastern Time on a business day. Typically, NASDAQ operates around 253 days a year.

\subsection{\textit{Ordeerbook Reconstruction}}

\vspace{5mm}
\noindent
The order book dynamics are simulated by continuously updating known information about its state. For example, given the previous day state of LOB for a stock, each new TotalView-ITCH event message changes the state of the book according to whether the message was an addition, update, cancellation or execution of the limit order. Figure 14 illustrates the incoming message “A” to add a limit order.  In response, a new order is added to the pool and the LOB is updated at the corresponding price level. In contrast, when an update message arrives – for example, cancellation “X” and “D”, the original order is found in the pool using order ID and updated accordingly.

\vspace{5mm}
\begin{figure}[hbt!]
    \centering
    \includegraphics[height = 8.5cm, width=10.5cm]{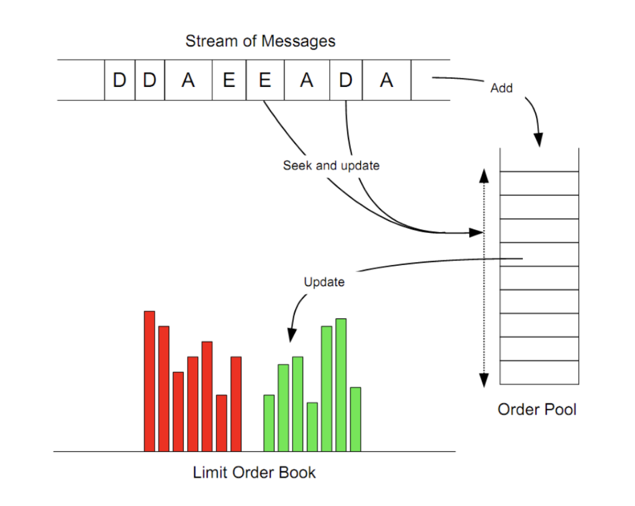}
    \caption{LOB reconstruction from NASDAQ data feed (HUANG, Ruihong and Polak, Tomas, 2011)}
    \label{fig:lobster_LOB}
\end{figure}

\vspace{5mm}
\noindent
For each LOBSTER data request, the LOB is reconstructed fully from the exchange message data for the stock during the specified period. By choosing an appropriate level of depth, only the number of levels requested is saved to the output file.

\vspace{5mm}
\noindent
LOBSTER data has been widely popular among academics in recent years. We find studies of Bouchaud et al. (2018), Cartea, Jaimungal and Penalva (2015), Bonart and Gould (2017) and others using LOBSTER data for microstructure research. The pre-processed output LOB files are available to download via the Internet saving a lot of time and effort, thus enabling academics to focus on the economic study.

\subsection{\textit{Data Prepossessing}}

\vspace{5mm}
\noindent
\textbf{Python Data Analysis}

\vspace{5mm}
\noindent
Once the data is downloaded from LOBSTER in CSV format, we define and deploy Python script to analyse and clean the data. Since each record in LOBSTER message file corresponds to exactly one record in orderbook file (and both files are ordered), we merge the data from messages and orderbook into one Python DataFrame object to be able to manipulate and analyse the data more efficiently. DataFrame is a two-dimensional tabular data structure with labelled rows and columns that can store data of various types - integer numbers, floating-point numbers, string characters and other. Throughout our empirical study we operate on market data using Python, and more specifically Python Pandas library which contains an extensive number of data science methods and functions implemented and available freely for use by academics and practitioners. Python has become very popular and is increasingly recognised as the de-facto data science language among both in recent years because of it’s easy to read (thus use) syntax and multi-paradigm nature. 

\vspace{5mm}
\noindent
\textbf{Inspection and Cleaning}

\vspace{5mm}
\noindent
To begin with, we investigate the properties of typical daily data. Our research objective is to define how various environment features, such as trade volume, influence the parameter of interest – market impact. To achieve this, we estimate the daily impact of MOs, and then calculate an average price impact for the total 6 months set of observations (amounting to 248 trading days). 

When looking at the average daily data, we discover that a large portion of events during trading hours constitute submission and cancellation of LOs (events of type 1 and 3 respectively). These findings, illustrated in Figure 15, are aligned with previous microstructure research (see Bouchaud et al. (2018)). Therefore, given our particular interest in how executions of MOs influence the price, only a small fraction of a daily dataset is relevant for the purposes of our quantitative analysis. To compensate for this, the study examines daily dynamics throughout a sufficiently long period of 6 months.
\clearpage
\vspace{5mm}
\begin{figure}[hbt!]
    \centering
    \includegraphics[height = 8.5cm, width=10.5cm]{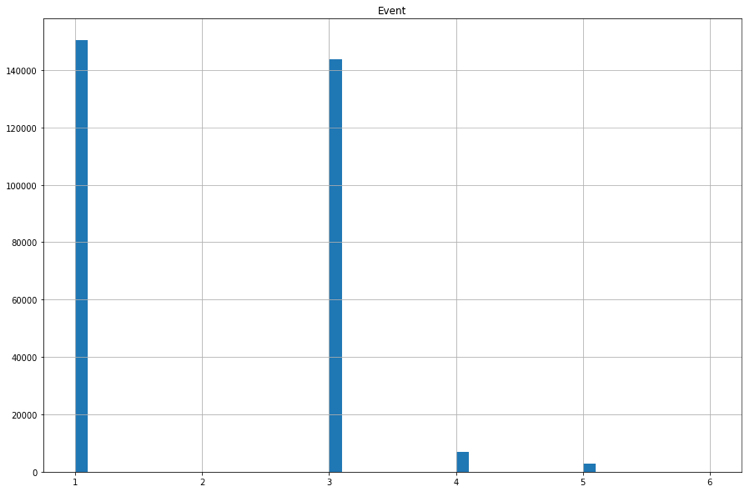}
    \caption{Frequencies of market events of each type during trading day.}
    \label{fig:event}
\end{figure}

\vspace{5mm}
\noindent
Next, when we examine the average trade volume for TSLA, some very unusual values are evident (Figure 16). Most electronic markets are known to operate during set time intervals throughout the day, where the beginning and end of the day are more generally reserved for auctions that result in the atypical trade activity. This might explain our findings of trade volume outliers.  

\clearpage
\begin{figure}[hbt!]
    \centering
    \includegraphics[height = 5cm, width=7cm]{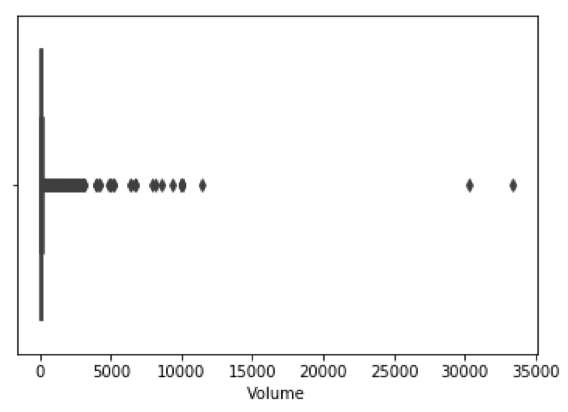}
    \caption{Box plot of trade volumes during the full trading day for TSLA stock}
    \label{fig:boxPlot1}
\end{figure}

\noindent
For each trading day, we remove the market activity for the first and the last hour, e.g. only using data for 10:30-15:00. This eliminates some inconsistency noticed in the previous examination of trade volume range as per Figure 17.

\begin{figure}[hbt!]
    \centering
    \includegraphics[height = 5cm, width=7cm]{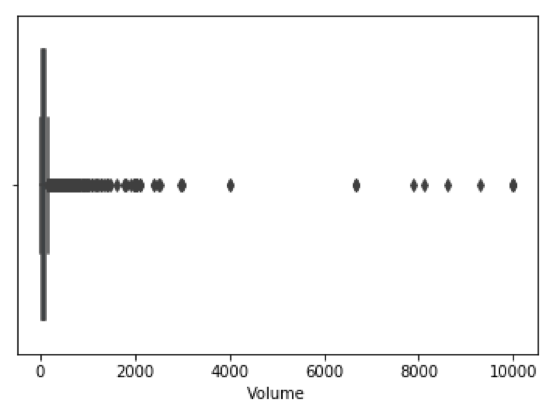}
    \caption{Box plot of trade volumes during 10:30-15:00 for TSLA stock}
    \label{fig:boxPlot2}
\end{figure}

\noindent
At this point, it is important to consider that whenever a single MO matches several LOs in the book, each LO execution is listed as a separate record in both LOBSTER messages and orderbook files, albeit with the same timestamp. An example is illustrated in Table 9 where three consecutive executions of visible LOs (event type 4) have the same timestamp, therefore, represent a single MO being matched with several LOs at that price. Interestingly, the execution of the last LO at the best ask price consumes all available liquidity at that level (33 shares) thus moving the ask price to the next best level – from 2058500 to 2058800. The next arriving MO to buy is most likely to be executed at this new price. This is an example of market impact, therefore, in our study, it is necessary to recognise that such multiple executions with the same timestamp correspond to a single MO. And the fact that it is the aggregated MO size that may have caused the price change, not just the latest matched LO that absorbed liquidity at the best quote. 

\clearpage
\begin{sidewaystable}
\centering
\begin{tabular}{>{\raggedright}p{2.2cm}lc c lc lc c lc c c c}
\toprule \toprule

Timestamp &  & Event Type   &  & OrderID         &  & Size    &  & Direction     &  & Ask Price            &  & Ask Volume & \\
\midrule

37837.0474 &  & 1           &  & 70920403        &  & 100       &  & \textit{-1}         &  & 2058500                 &  & 100 & \\

\textbf{37837.0474}  &  &   \textbf{4}        &  & \textbf{70920403}   &  & \textbf{22}    &  & \textbf{-1}    &  & \textbf{2058500}    &  & \textbf{78} & \\
\textbf{37837.0474}  &  &   \textbf{4}        &  & \textbf{70920403}   &  & \textbf{45}    &  & \textbf{-1}    &  & \textbf{2058500}    &  & \textbf{33} & \\
\textbf{37837.0474}  &  &   \textbf{4}        &  & \textbf{70920403}   &  & \textit{33}    &  & \textbf{-1}    &  & \textit{2058800}    &  & \textit{69} & \\
\textbf{37837.0479}  &  &   \textbf{3}        &  & \textbf{70920380}   &  & \textbf{100}   &  & \textbf{1}     &  & \textbf{2058800}    &  & \textbf{69} & \\

37837.0479 &  & 1           &  & 70920420        &  & 100       &  & 1          &  & 2058800                 &  & 69 & \\

\bottomrule

\end{tabular}
\label{table:loExecution}
\caption{Separate LO executions representing \\ the same MO at the same timestamp}
\end{sidewaystable}

\clearpage
\noindent
Consequently, we group LO executions that happened at exactly the same time, and aggregate individual LO volumes to reconstruct the size of the original MO.

\subsection{\textit{Machine Learning}}

\vspace{5mm}
\noindent
\textbf{Unsupervised Learning}

\vspace{5mm}
\noindent
Unsupervised learning describes a set of algorithms that do not make use of labelled responses associated with the data. They alternatively exploit the underlying data structure to draw inference. In this respect, there are no correct target variables, instead, we are only given inputs, $\mathcal{D}={\{x_i\}}_{i=1}^N$, to deduce meaningful properties of the probability density of the dataset (given no training sample). This is sometimes referred to as \textit{knowledge} discovery and is a much less well-defined problem as there is no means of evaluating the learners' performance given no obvious target metric. 

\vspace{5mm}
\noindent
\textbf{Reinforcement Learning}

\vspace{5mm}
\noindent
Reinforcement learning constitutes the third class of ML. Reinforcement Learning (RL) attempts to learn a policy of how to take actions in a dynamic environment to maximise the discounted future reward criterion. To collect information the learner agent actively interacts with the environment (exploring unknown actions to gain new insight and exploiting the information already collated) and in some cases affects the system.

\vspace{5mm}
\noindent
\textbf{Parametric and Non-Parametric Models}

\vspace{5mm}
\noindent
\begin{itemize}
    \item 	Parametric statistical learning involves a specified model of form $f(y)$ with fixed a fixed number of parameters that define its behaviour. The canonical example of a parametric model is that of linear regression. This involves estimation of a set of $p+1$ coefficients relating to the vector $\beta=(\beta_0,\beta_1,\ldots\beta_p)$ whereby a response $y$ is linearly proportional to each feature $x_{j}$. Parametric models hold the advantage of being efficient in use, but the disadvantage of making strong assumptions about the nature of the distribution of the dataset.  
    
    \item 	An alternative is to consider a form for $f$ where the number of parameters grows with respect to the size of the dataset. Non-parametric models may not involve any parameters or may describe a distribution with a finite number of parameters. These are more flexible but often computationally intractable given the large quantity of observational training data required to account for the lack of assumption. All models considered in this research fall under the parametric classification. As always, by increasing the number parametric vectors in our feature space we risk the danger of overfitting the model as the model may follow the \textit{“noise”} too closely as opposed to the \textit{“signal”}.

\end{itemize}

\vspace{5mm}
\subsection{\textit{The Gauss–Markov Theorem}}

\vspace{5mm}
\noindent
To assess the statistical significance of the model and conduct inference, we make assumptions about the residuals – properties of the unexplained aspect of the input sample. The \textit{Gauss–Markov theorem} (GMT) defines several assumptions required for OLS to generate unbiased estimates of the model parameters $\beta$. Under the standard GMT, the coefficients’ best (optimal) linear unbiased estimator is given by OLS if and only if:

\begin{enumerate}

    \item 	The data for the input variables $x,\ldots,x_k$ is a random sample from the population
  
    \item 	The errors have a conditional mean of zero given any of the inputs:\\ $\mathbb{E}[\epsilon|x,\ldots,x_k] = 0$
    
    \item Homoskedasticity, the error term $\epsilon$ has a constant variance given the inputs:\break $\mathbb{E}[\epsilon|x,\ldots,x_k] $ = $\sigma^2$

\end{enumerate}

\noindent
The best estimator is the one that yields the lowest standard error (variance) of the estimates. In addition to these GMT assumptions, the classical linear model assumes \textit{normality} – the population error is normally distributed and independent of the input variables. This implies that the output variables are normally distributed (conditional on the input variables) permitting the derivation of the exact distribution of coefficients.

\vspace{5mm}
\subsection{\textit{Statistical Inference}}

\vspace{5mm}
\noindent
The functional relationship produced by supervised learning algorithms can be used for \textit{inference} (that is, to gain new insights into how the outcomes are generated) or for \textit{prediction} (that is, to generate accurate outcome estimates, represented by $\hat{y}$) for unknown or future inputs ($x$).

\vspace{5mm}
\noindent
In the context of regression, inference looks to draw conclusions about the true relationship of the \textit{population regression function} (PRF) using the sample observations. This means running tests of hypothesis about the significance of the overall relationship estimated by SRF, as well as testing for the significance of particular coefficients. We also estimate confidence intervals for the estimated model.

\vspace{5mm}
\noindent
An important component of statistical inference is a \textit{test statistic} with a known distribution. This assumes, under the \textit{null hypothesis}, the estimated statistic is similar to the true model; and computes the probability (given by the $p$-value) of observing the value for this statistic in the sample. If the p-value drops below a confidence threshold – usually five percent – we reject the null hypothesis. 

\vspace{5mm}
\noindent
In addition to GMT assumptions, we have the following distributional characteristics for test statistic exactly when normality holds: The test statistic for a hypothesis test given an individual coefficient $\beta_j$ is 

\begin{eqnarray}
  t_j =  \frac{\hat{\beta_j}}{\hat{\sigma} \sqrt{v_j}} \sim  t_N-p-1
\end{eqnarray}

\vspace{5mm}
\noindent
and follows a \textit{t distribution} of $N-p-1$ degrees of freedom, where $v_j$ is the $j^{th}$ element of the diagonal of $(x^T x)^{-1}$.

\vspace{5mm}
\subsection{\textit{AIC and BIC Goodness-of-Fit Measures }}

\vspace{5mm}
\noindent
Alternative prominent goodness-of-fit measures are the \textit{Akaike Information Criterion} and (AIC) and the \textit{Bayesian Information Criterion} (BIC). The goal is to minimise these based on the MLE:

\begin{itemize}
  \item 	AIC = $-2log(\mathcal{L}^\ast)+2k$, where $\mathcal{L}^\ast$ denotes the value of the maximised likelihood function and $k$ is the number of parameters
  
  \item 	BIC = $2log(\mathcal{L}^\ast)+\ log(N)k$, where $N$ is the sample size
  
\end{itemize}

\vspace{5mm}
\noindent
Given a collection of estimated models, AIC determines the quality of a model by looking to find the model that best describes an unknown data-generating process; whereas BIC aims to find the best model amongst a finite set of candidates. Both provide a means of model selection. Each metric penalises for complexity. While AIC has a lower penalty, thus might overfit; BIC imposes a larger penalty so it can underfit the data. In practice, AIC and BIC can be used jointly to guide the process of model selection if the objective is to fit in-sample; otherwise, \textit{cross-validation} is preferable.

\vspace{5mm}

\subsection{\textit{Hidden Orders}}

\vspace{5mm}
\noindent
The LOBSTER data provides information on hidden LO executions – market event of type 5. When reconstructing a MO that had been matched against several LOs in the book, we often observe a MO initially execute against \textit{hidden} LO, only matching resting \textit{visible} LOs at the best quote once it has consumed all hidden liquidity. Although we do not particularly speak of the effect of hidden orders in this text, it is important to understand that the response (effect) of some of the visible MOs is affected by the hidden LOs at a better price. From the earlier description of NASDAQ electronic exchange, we recall that orders are executed in specific successions: orders with the best prices are matched first, priority being given to the visible orders and the time of order submission. In other words, visible orders are always executed first unless hidden LO has a better price. In such cases, hidden orders (hidden liquidity) dilute the effect of aggressive MOs at the price that matches the best in the LOB.

\subsection{\textit{Empirical Studies Review}}

\vspace{5mm}
\noindent
In our implementation of price response algorithm we followed a number of prominent (in the field of financial microstructure) research publications that worked with LOBSTER data such as Bouchaud et al. (2018) and Gould et al. (2013). Using the data for the same stocks, our results underestimate $\mathcal{R}\left(1\right)$ for all four observed tickers: SIRI, EBAY, TSLA, PCLN. Since our findings report average spread $\langle \textbf{s} \rangle$ exactly matching results of BBouchaud et al. (2018), we assume that the discrepancy in the calculated values of response function $\mathcal{R}\left(1\right)$ arises from different approaches in dealing with a MO that matched a number of LOs in the LOB (therefore appears as multiple entries at the same time-stamp in the LOBSTER message file).

\clearpage
\begin{sidewaystable}
\centering
\small
\begin{tabular}{>{\raggedright}p{2.6cm}lc c lc lc c lc c c c c c}
\toprule \toprule

                      && nobs            && min             && max          && mean         && variance          && skewness               && kurtosis & \\  
                    \midrule

Order-Flow Imbalance      &&253           && -2718.71       &&2330.0       &&-191.11        &&2142316.26         &&7.64            &&-1.20 \\ \cmidrule(r){1-1}

Aggregated Market Impact  &&253           && -21.625       &&19.33       &&0.58          &&62.91          &&-0.05             &&-1.31 \\

\bottomrule

\end{tabular}
\label{table:descriptive_Stats1}
\caption{Descriptive Statistics: Order-Flow Imbalance and Aggregated Market Impact. Order-Flow Imbalance measured in shared, Aggregated Market Impact is measured in dollar cents.}
\end{sidewaystable}

\clearpage
\begin{sidewaystable}
\centering
\small
\begin{tabular}{>{\raggedright}p{2.4cm}lc c c lc c lc c c c c c c}
\toprule \toprule

            &&            &             & Linear Regression          &         &          &               && \\\\  
\midrule

MSE      &&4.51           &4.42       &4.75       &5.03        &3.87         &4.64            &4.26    &3.72  &4.08  &4.97 \\ \cmidrule(r){1-1}

Mean    &&4.42  \\
\cmidrule(r){1-1}

St.Dev  &&0.42   \\\\\\

\midrule\midrule 

            &&            &             & Decision Tree               &         &          &               && \\\\

MSE      &&2.91           &4.77       &6.00       &4.08        &2.47        &3.98            &4.51        &2.94      &3.91  &5.44  \\ \cmidrule(r){1-1}

Mean     &&4.11 \\
\cmidrule(r){1-1}

St.Dev  &&1.07  \\\\

\bottomrule

\end{tabular}
\label{table:descriptive_Stats2}
\caption{MSE Cross Validation Scores run $k \times 10$ . MSE is measured in dollar cents.}
\end{sidewaystable}

\clearpage
\section{}
\subsection{\textit{Future Work}}

\vspace{5mm}
\noindent
\subsubsection{Alternative Data}

\vspace{5mm}
\noindent
When giving a functional definition of market impact we have emphasised that in this work we are looking specifically at the impact of individual market orders. This is partially due to the specifics of LOBSTER data set that does not identify meta-orders. However, as we later discovered, there are other free resources such as NASDAQ OMX NORDIC (or just OMX) online database where one can access trades and quotes data that identify market participants. This would enable reconstruction of meta-orders and thus study of the latent liquidity. 

\vspace{5mm}
\noindent
Alternatively, as suggested by many similar empirical studies, having proprietary data makes a big difference for accurate research of financial markets microstructure.  Previous empirical studies of market impact employed traditional statistical techniques with a few publications such as Nevmyvaka and Kearns (2016) and Lehalle (2018) discussing application of ML methods. The majority of such studies were conducted on clean academic historical datasets that often omit microstructure nuances required for a deeper level investigation.
Contrastingly, we seek collaboration with industry partners to support the vision of data driven research, by providing real-time proprietary microstructure data. This can be used to simulate interesting game theory variations of agents behaviours in varying market conditions.

\vspace{5mm}
\noindent
\subsubsection{Illiquid Markets}

\vspace{5mm}
\noindent
One of the main distinguishing properties of different markets is the availability and visibility of liquidity. Moreover, liquidity can change drastically for an asset class in times of stressed markets. Such events, characterised as \textit{black swan} events, are challenging to model since they appear as outliers in the data, making it difficult for practitioners to decide on the most optimal actions when facing such scenarios. The second motivation for this research is to investigate the key drivers affecting liquidity dynamics, under stressed market conditions.

\vspace{5mm}
\noindent
\subsubsection{Invariant Market Impact Function}

\vspace{5mm}
\noindent
Kyle and Obizhaeva (2018), as well as, Eisler and Bouchaud (2016) have researched into a universal solution for price impact. In their paper, \textit{“The Market Impact Puzzle”}, the former propose two conditions on the drivers of invariant formula which result in the tightly parameterized function. Eisler and Bouchaud (2016) have successfully applied the propagator technique to estimation of price impact in OTC credit index market, with their quantitative results being similar to those in more traditional asset classes. Thus, confirming that price impact is a universal phenomenon regardless of specifications of market microstructure.

\vspace{5mm}
\noindent
\subsubsection{Optimal Execution}

\vspace{5mm}
\noindent
Having a more comprehensive understanding of liquidity (ability to forecast liquidity more accurately) is imperative to measuring market impact of trades, hence associated transaction costs. This knowledge can help optimize a liquidation schedule for large block orders. Studies of optimal trading strategies have often solely focused on a single agent who desires to trade a certain amount of the security. When generalised to multiple agents, the resulting stochastic optimal control problem is extremely difficult to solve. One approach is to investigate a mean field game framework in which the interactions of major and minor players are approximated by analysing how a finite subsample $\mathcal{N}$,\ of the population of agents who seek to adopt optimal game strategy, are restricted in their behaviour as $\mathcal{N}\longrightarrow\infty$. Such situations constitute Nash Equilibrium. 

\vspace{4mm}
\noindent
In this context, an area of machine learning called Reinforcement Learning (which has roots in control theory) can be employed to solve the problem of optimised trade execution. Reinforcement Learning (RL) attempts to learn a policy of how to take actions in the environment in order to maximise the discounted future reward criterion.  The key advantage of Reinforcement Learning over other Machine Learning techniques, is that the agent learns directly how to make decisions, as opposed to predicting target values. In the framework of our stochastic control problem, the agent learns to optimize a trading path by dividing a specified order across time, and venues with potentially different liquidity profiles.

\vspace{4mm}
\noindent
To supplement the learning process, we will look to explore further applications of Deep Neural Networks. As the agent accumulates the knowledge of which actions yield the highest reward, this information is continuously stored to assist future decision making ( trade-off between \textit{exploration} and \textit{exploitation}). However, as the learning process continues this knowledge space can become extremely large making it unfeasible for the algorithm to process (both mathematically and computationally intractable for partially-observable/ stochastic environments). An alternative to stowing all previously encountered states and their corresponding best actions is to generalise past experiences by creating a neural network to predict the reward for a given input. This approach is more tractable than standard knowledge storing techniques, allowing RL to be applied to more complex problems where many network layers can capture the most intricate details, thus facilitating deep learning.

\vspace{5mm}
\noindent
To affirm the validity of our findings, a series of experiments will be carried out using the scientific method. The work will be conducted in a controlled environment with initial offline verification of assumptions on historical data sets. Some properties of real time market dynamics, however, cannot be verified offline; therefore, should be validated at runtime. This might require high-performance computational resources. The success of derived optimal trading behaviour will be benchmark against traditional definitions of optimal execution such as arrival price, VWAP and participation percentage.

\vspace{5mm}
\subsubsection{Computational Resources}
\vspace{5mm}
\noindent
High frequency and detailed precision of LOBSTER data imply that the amount of information about certain stocks' trade activity easily exceeds several of Giga Bites (GBs). Such a high volume of data is best dealt with using high-performance computing resources that are optimised to improve on processing timings. 

\vspace{5mm}
\noindent
We strongly believe that the pace of learning from data can be majorly improved by the usage of appropriate hardware designed for efficient analysis of large quantities of information.

\vspace{5mm}
\noindent
The below chart illustrates the average computational time for estimation of the lag-1 unconditional impact of trades for four selected stocks. One of the challenges of the current study was the long time it took to test the algorithm and collect statistical information for further inference.   

\clearpage
\begin{figure}[hbt!]
    \centering
    \includegraphics[height = 6.5cm, width=8.5cm]{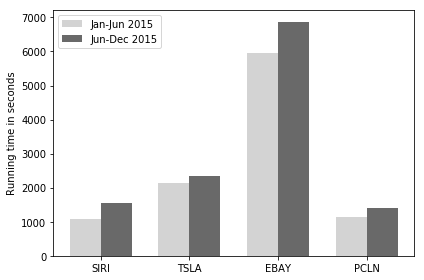}
    \caption{Computational time in seconds for estimation of lag-1 unconditional impact for $2^{nd}$ of January to $30^{th}$ of June 2015 in light grey; and the computational time in seconds for estimation of lag-1 unconditional impact $30^{th}$ of June to $31^{st}$ of December 2015 in dark grey}
    \label{fig:compPerformance}
\end{figure}

\vspace{5mm}
\noindent
It is evident that even for less actively traded stocks such as SIRI and PCLN an average running time of an algorithm that estimates lag-1 unconditional market impact for 6 months is above 1000 seconds (the equivalent of 15 min). For a stock that is traded heavily on NASDAQ, such as EBAY, estimation of the average response function for half a year worth of data can easily take over an hour. 

\section{}

\end{appendices}

\end{document}